\documentclass{article} %
\usepackage{iclr2026_conference,times}

\usepackage{amsmath,amsfonts,bm}

\def\eqref#1{equation~\ref{#1}}

\def\1{\bm{1}}

\DeclareMathAlphabet{\mathsfit}{\encodingdefault}{\sfdefault}{m}{sl}
\SetMathAlphabet{\mathsfit}{bold}{\encodingdefault}{\sfdefault}{bx}{n}

\usepackage{hyperref}
\usepackage{url}
\usepackage{graphicx}
\usepackage{booktabs}
\usepackage{subcaption}
\usepackage[ruled]{algorithm2e} %
\usepackage[T1]{fontenc}  %

\usepackage{multirow}

\usepackage{geometry}

\usepackage{amssymb}
\usepackage{pifont}
\usepackage{booktabs}
\usepackage{array}

\newcommand{\cmark}{\checkmark}%
\newcommand{\xmark}{\ding{55}}%
\title{Generative Blocks World: \\Moving Things Around in Pictures}

\author{Vaibhav Vavilala$^{1}$~~~~ Seemandhar Jain$^{1}$~~~~ Rahul Vasanth$^{1}$~~~~ D.A. Forsyth$^{1}$~~~~ Anand Bhattad$^{2}$ \\\vspace{-2mm}
\\
$^{1}$University of Illinois Urbana-Champaign~~~~$^{2}$Johns Hopkins University \\
}

\iclrfinalcopy %
\begin{document}

\makeatletter
\g@addto@macro\@maketitle{
	\begin{figure}[h]
	\scriptsize
	\setlength{\linewidth}{\textwidth}
		\setlength{\hsize}{\textwidth}
  \centering
  \footnotesize
  \setlength\tabcolsep{0.2pt}
  \renewcommand{\arraystretch}{0.1}
  		\vspace{-7mm}
  \includegraphics[width=1.0\textwidth]{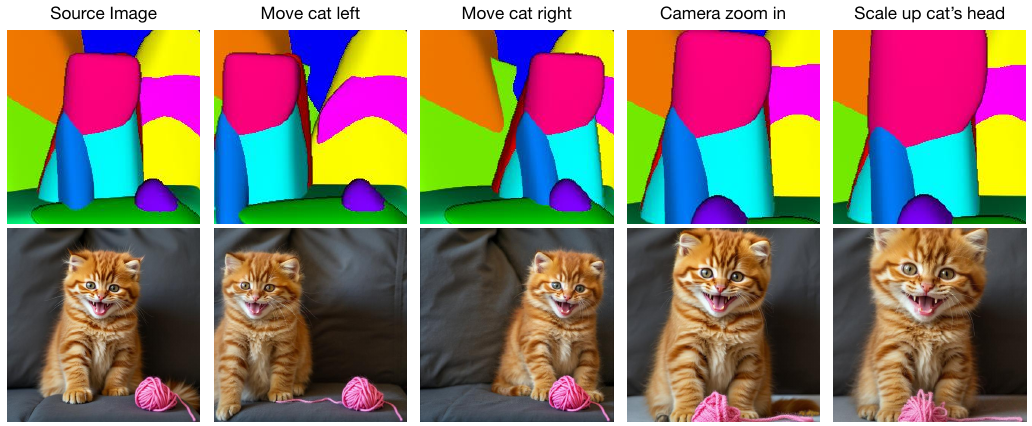}
  \vspace{-12pt}
  \caption{\textbf{Generative Blocks World.} Given an input image (bottom left), we extract a set of 3D convex primitives (top left) that provide an editable and controllable representation of the scene. These primitives are used to generate new images that respect geometry, texture, and the text prompt. The first column shows the original input and its primitive decomposition. Subsequent columns show sequential edits: translating the cat to the left (second column), translating it to the right (third column), moving the yarn in front of the cat and shifting the camera toward the scene center (fourth column), and scaling up the cat’s head (burgundy primitive; fifth column). Our method enables 3D-aware semantic image editing through intuitive manipulation of these learned primitives.}
  \label{fig:teaser}
\end{figure}
}
\maketitle

\begin{abstract}
We describe Generative Blocks World to interact with the scene of a generated image by manipulating simple geometric abstractions. Our method represents scenes as assemblies of convex 3D primitives, and the same scene can be represented by different numbers of primitives, allowing an editor to move either whole structures or small details. Once the scene geometry has been edited, the image is generated by a flow-based method, which is conditioned on depth and a texture hint. Our texture hint takes into account the modified 3D primitives, exceeding the texture-consistency provided by existing techniques. These texture hints (a) allow accurate object and camera moves and (b) preserve the identity of objects. Our experiments demonstrate that our approach outperforms prior works in visual fidelity, editability, and compositional generalization.
\end{abstract}

\section{Introduction}
\label{sec:intro}
There is a rich literature treating editing real and generated images using various image-centered interfaces like dragging features. %
Interaction paradigms that exploit explicit representations of 3D are much less common.
This paper describes an image editor built on a full 3D interaction paradigm,
using a representation that is both compact and accurate -- a {\em generative blocks world}.

\begin{figure}[t!]
  \centering
  \vspace{-2pt}
  \includegraphics[width=1.0\textwidth]{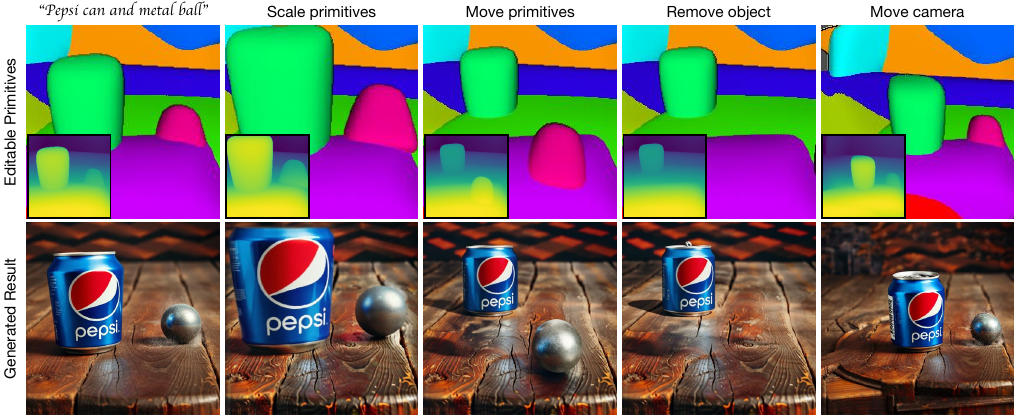}
  \vspace{-12pt}
  \caption{\textbf{Editable Primitives as a Structured Depth Prior for Generative Models.} Our method uses 3D convex primitives as an editable intermediate representation from which depth maps are derived. These depth maps (shown as insets in the top row) are used to condition a pretrained depth-to-image generative model. The top row shows primitive configurations after sequential edits—translation, scaling, deletion, and camera motion—alongside their corresponding derived depth maps. The bottom row shows the resulting synthesized images. Unlike direct depth editing, which is unintuitive and underconstrained, manipulating primitives offers a structured, interpretable, and geometry-aware interface for controllable image generation.
  }
  \label{fig:teaser_2}
      \vspace{-3pt}
\end{figure}

Any scene representation that supports a {\bf camera move} has some form
of 3D representation.  An explicit 3D representation helps, Fig~\ref{fig:teaser}. %
Explicit 3D representations have other important advantages.  First, they offer {\bf shape constancy}.
When an object is moved across a perspective view, it is seen from a new aspect
because the location of the focal point moves in object coordinates.  This means that (a) the shape of the object may change,
with a change that depends on the field of view of the camera and the shape of the object 
and (b) some surface markings will become visible or invisible (e.g. bar code on soda can, Fig~\ref{fig:teaser_2}).
When an object is moved toward or away from the camera, its image should expand or shrink (e.g. cat, Fig~\ref{fig:teaser}).  If an editor does
not preserve these properties correctly, the viewer may conclude that the shape or size of the object has changed.
A properly constructed 3D representation will prevent this.  Second, they offer {\bf contact consistency}.  A user who moves (say) a tin on a table
generally expects the tin to remain in contact with the table.  An explicit 3D representation allows the user to manage whether it does or not (e.g. dog in Fig~\ref{fig:scale_1}; soda can, Fig~\ref{fig:teaser_2}).
Third, they offer {\bf shape completion}.  Objects have backs that are not visible, but may have an effect when another object is moved in a scene.
An explicit 3D representation can capture this effect.

It has been hard to build a 3D representation that: (a) represents the scene accurately enough that edited images are realistic and
(b) is compact enough to support interactions.    This paper uses modern fitting methods to represent
scenes as small assemblies of meaningful parts or primitives (cf. \emph{Blocks World}~\cite{roberts} or \emph{geons}~\citet{biederman}). We call our
method \textbf{Generative Blocks World}, \emph{though our learned primitives are richer than cuboids}. 
Our method yields assemblies by decomposing an input image
into a sparse set of convex polytopes~\citep{vavilala2024improvedconvexdecompositionensembling} that approximate the scene’s depth map well enough to enable view-consistent texture projection.  Further,
our primitives respect object boundaries rather well. A user can reach into the scene and move a primitive, with predictable results.
Our simple scene representations yield {\em hints} as to the appearance of the final image.  These hints, together with the
primitive depth map, are inputs to an off-the-shelf image generator, which renders accurate images.

Primitive decompositions have very attractive properties.
They are \emph{selectable}: individual primitives can be intuitively selected and manipulated (Fig.~\ref{fig:teaser}). They are
\emph{object-linked}:  a segmentation by primitives is close to a
segmentation by objects, meaning an editor is often able to move an
object or part by moving a primitive (Figs~\ref{fig:teaser}; \ref{fig:teaser_2}; \ref{fig:edit_1}). They are \emph{accurate}:  the depth map from a properly
constructed primitive representation can be very close to the original
depth map (Section~\ref{sec:decomp}), which means primitives can be
used to build texture hints (Section~\ref{sec:method_diffusion}) that
support accurate camera moves (Figs~\ref{fig:method}; \ref{fig:cammovegood}).  
They have \emph{variable scale}:  one can represent the same scene
with different numbers of primitives, allowing multi-scale edits (Figs~\ref{fig:scale_1}; \ref{fig:scale_2}; \ref{fig:PartAbl1}).

{\bf Contributions:} \textbf{(1)} We describe a pipeline that fuses convex primitive abstractions with a flow-based generator to yield a natural 3D interaction paradigm for image editing. Our pipeline uses a texture‑hint procedure that supports camera moves and edits at the object-level, while preserving identity. \textbf{(2)} We provide extensive evaluation demonstrating superior geometric control, texture retention, and edit flexibility relative to recent state-of-the-art baselines.

\section{Related Work}
\label{related}
{\bf Primitive Decomposition:} Early vision and graphics pursued parsimonious part-based descriptions, from Roberts’ \emph{Blocks World}~\cite{roberts} and Binford’s generalized cylinders~\cite{binford71} to Biederman’s geons~\cite{biederman}. Efforts to apply similar reasoning to real-world imagery have been periodically revisited~\cite{GuptaEfrosHebert_ECCV10, monnier2023dbw, bhattad2025visualjengadiscoveringobject} from various contexts and applications. Modern neural models revive this idea: BSP-Net~\cite{Chen2019BSPNetGC}, CSG-Net~\cite{Sharma_2018_CVPR}, and CVXNet~\cite{deng2020cvxnet} represent shapes as unions of convex polytopes, while Neural Parts~\cite{abstractionTulsiani17}, SPD~\cite{Zou_2018_CVPR}, and subsequent works~\cite{liu2022towards} learn adaptive primitive sets. Recent systems extend from objects to scenes: Convex Decomposition of Indoor Scenes (CDIS)~\cite{Vavilala_2023_ICCV} and its ensembling/Boolean refinement~\cite{vavilala2024improvedconvexdecompositionensembling} fit CVXNet-like polytopes to RGB-D images, using a hybrid strategy. CubeDiff~\cite{kalischek2025cubediffrepurposingdiffusionbasedimage} fits panoramas inside cuboids. Our work leverages CDIS as the backbone, but (i) improves robustness to in-the-wild depth/pose noise and (ii) couples the primitives to a Rectified Flow (RF) renderer, enabling controllable synthesis.

{\bf Conditioned Image Synthesis:} Layout-to-image translation was pioneered in GANs~\cite{isola2017image, zhu2017unpaired, park2019spade} and is now dominated by diffusion models such as Stable Diffusion~\cite{rombach2021highresolution}, ControlNet~\cite{zhang2023adding}, and T2I-Adapter~\cite{mou2023t2i}. These models can compose multiple spatial controls~\citep{vavilala2024denoisingmontecarlorenders}, perform color edits~\citep{vavilala2024dequantizationcolortransferdiffusion} and relight scenes~\cite{xing2025luminet}. We utilize a pretrained depth-conditional FLUX model, conditioning it on depth maps derived from our 3D primitives.

{\bf Point-Based Interactive Manipulation:}   Methods like DragGAN~\citep{pan2023drag} and its diffusion-based successors~\citep{shi2024dragdiffusion, mou2023dragondiffusion, cui2024stabledrag, 10657993} offer intuitive 2D control by dragging handle points. Some approaches extend this to 3D using NeRFs for multi-view consistency~\cite{Guang_2025_CVPR} or leverage self-guidance for layout control~\citep{epstein2023selfguidance}. Our work differs by operating on editable 3D primitives instead of 2D points. This enables multi-resolution control and camera movement while handling perspective, occlusion, and texture.

{\bf Object-Level and Scene-Level Editing:} Many recent works embed 3D priors for editing, though often focusing on single objects~\cite{gu2021stylenerf, wang2022score, poole2022dreamfusion, tang2023make, cheng20253d} or using language to guide transformations~\cite{michel2023object}. Our Generative Blocks World generalizes to complex edits not easily described by text. Another paradigm, seen in Image Sculpting~\cite{yenphraphai2024image} and OMG3D~\cite{zhao20253d}, reconstructs an explicit 3D mesh for manipulation before re-rendering. While precise, these multi-stage pipelines can be bottlenecked by reconstruction quality. Our method provides a more streamlined approach by operating on abstract primitives, achieving strong geometric control without the complexity of direct mesh manipulation.

{\bf Primitive-Based Scene Authoring:} LooseControl~\citep{LooseC} enables control via box-like primitives by fine-tuning a diffusion model with LoRA weights. This training is necessary to bridge the domain gap between its coarse primitive-based depth and standard depth maps~\citep{depth_anything_v2}. In contrast, our underlying primitive representation is accurate enough to require no fine-tuning. Furthermore, by abstracting objects into single, monolithic boxes, LooseControl is limited to holistic transformations and cannot perform part-level edits. Our method uses structured geometry, decomposing objects into multiple convex polytopes at variable levels of detail for more granular control. A more recent work, Build-A-Scene~\citep{eldesokey2024build}, uses a similar pipeline to LooseControl and thus inherits its limitations. Our approach differs by: (i) decomposing objects into multiple convex polytopes for finer control, (ii) supporting camera movement, and (iii) allowing novel scenes to be authored from scratch via primitive assembly.

\section{Method}
\label{sec:method}

\begin{figure*}[t!]
\vspace{-2pt}
\centering
\includegraphics[width=\textwidth]{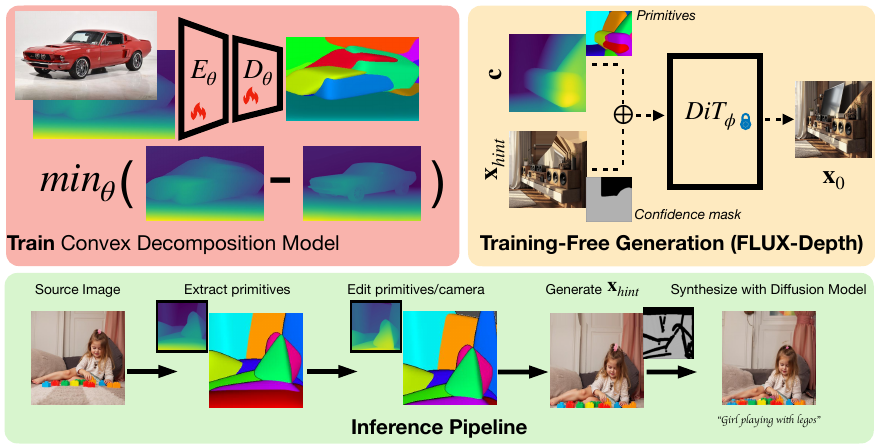}
\vspace{-15pt}
\caption{\textbf{Pipeline Overview.} \textbf{Top left:} We use convex decomposition models~\cite{vavilala2024improvedconvexdecompositionensembling} to extract primitives from an input image at multiple scales. \textbf{Bottom:} Users can manipulate these primitives and the camera to define a new scene layout. We render the modified primitives into a depth map and generate a texture hint image. These serve as inputs to a pretrained depth-to-image model~\cite{flux2024}, which requires no fine-tuning (\textbf{Top right}). The generated image respects the modified geometry, preserves texture where possible, and remains aligned with the text prompt.}
\label{fig:method}
\vspace{-12pt}
\end{figure*}

Generative Blocks World generates realistic images conditioned on a parsimonious and editable geometric representation of a scene: a set of convex primitives. The process consists of four main stages (Fig.~\ref{fig:method}): (i) primitive extraction from any image via convex decomposition (Sec.~\ref{sec:decomp}), 
(ii) generating an image conditioned on the primitives (and text prompt), (iii) user edits the primitives and/or camera, and (iv) generates a new image conditioned on the updated primitives, while preserving texture from the source image (Sec.~\ref{sec:texture}).  
We describe each component in detail below. 

\subsection{Convex Decomposition for Primitive Extraction}
\label{sec:decomp}
Our primitive vocabulary is blended 3D convex polytopes as described in~\cite{deng2020cvxnet}. CVXnet represents the union of convex polytopes using indicator functions $O(x) \rightarrow [0,1]$ that identify whether a query point $x \in \mathbb{R}^3$ is inside or outside the shape. Each convex polytope is defined by a collection of half-planes.

A half-plane $H_h(x) = n_h \cdot x + d_h$ provides the signed distance from point $x$ to the $h$-th plane, where $n_h$ is the normal vector and $d_h$ is the offset parameter.

While the signed distance function (SDF) of any convex object can be computed as the maximum of the SDFs of its constituent planes, CVXnet uses a differentiable approximation. To facilitate gradient learning, instead of the hard maximum, the smooth LogSumExp function is employed to define the approximate SDF, $\Phi(x)$:
\vspace{-2pt}
$$\Phi(x) = \text{LogSumExp}\{\delta H_h(x)\}
\vspace{-2pt}$$
The signed distance function is then converted to an indicator
function $C: \mathbb{R}^3 \rightarrow [0,1]$ using: $C(x|\beta) = \text{Sigmoid}(-\sigma\Phi(x)).$

The collection of hyperplane parameters for a primitive is denoted as $h = \{(n_h, d_h)\}$, and the overall set of parameters for a convex as $\beta = [h, \sigma]$. While $\sigma$ is treated as a hyperparameter, the remaining parameters are learnable. The parameter $\delta$ controls the smoothness of the generated convex polytope, while $\sigma$ controls the sharpness of the indicator function transition. The soft classification boundary created by the sigmoid function facilitates training through differentiable optimization. For our primitive model we use ResNet-18 Encoder $E_\theta$ followed by 3 fully-connected layers that decode into the parameters of the primitives $D_\theta$. While the model is lightweight, the SOTA of primitive prediction requires a different trained model for each primitive count $K$.

Recent work has adapted primitive decomposition to real-world scenes (as opposed to well-defined, isolated objects, such as those in ShapeNet~\cite{Vavilala_2023_ICCV}). These methods combine neural prediction with post-training refinement: an encoder-decoder network predicts an initial set of convex polytopes, which is followed by gradient-based optimization to align the primitives closely to observed geometry. This approach is viable because the primary supervision for primitive fitting is a depth map (with heuristics that create 3D samples, and auxiliary losses to avoid degenerate solutions). Note that ground truth primitive parameters are not available (as they could be in many other computer vision settings e.g., segmentation~\cite{Kirillov2023SegmentA}). This is why the losses encourage the primitives to classify points near the depth map boundary correctly instead of directly predicting the parameters.

\textbf{Rendering the primitives.} We condition the RF model on
the primitive representation via a depth map, obtained by ray-marching
the SDF from the original viewpoint of the scene.  Depth conditioning
abstracts away potential `chatter` in the primitive representation
from e.g. over-segmentation, while simultaneously yielding flexibility
in fine details (depth maps typically lack pixel-level
high-frequency details). Depth-conditioned image synthesis models
are well-established e.g.~\cite{zhang2023adding}.
Because \textbf{it's
  hard to edit a depth map, but easy to edit 3D
  primitives}, our work adds a new level of control to the existing image
synthesis models.  As we establish quantitatively in
Table~\ref{tab:absrel_vs_parts}, our primitive generator is extremely
accurate, and our evaluations show that we get very tight control over
the synthesized image via our primitives. This means that
whatever domain gap there is between depth from primitives and depth from SOTA depth estimation
networks is not significant.

\textbf{Scaling to in-the-wild scenes.}
We collect 1.8M images from LAION to train our primitive prediction models. To obtain ground truth depth supervision, we use DepthAnythingv2~\cite{depth_anything_v2}. We lift the depth map to a 3D point cloud using the pinhole camera model.

\subsection{Depth-Conditioned Inpainting in Rectified Flow Transformers}
\label{sec:method_diffusion}

\begin{figure}[t!]
\vspace{-3pt}
\centering
\includegraphics[width=\linewidth]{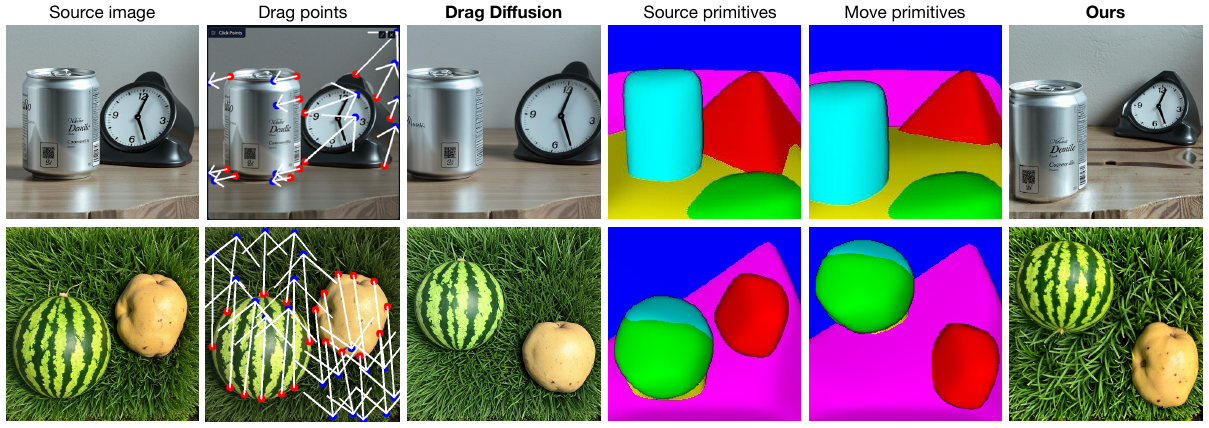}
\vspace{-15pt}
\caption{\textbf{Comparison with Drag Diffusion}~\citep{shi2024dragdiffusion}. \textbf{First row:} Given a scene (first column), we attempt to reposition objects using a recent point-based image editing method by drawing drag handles (second column). However, drag points are ambiguous: it is unclear whether the intended operation is translation or scaling. As a result, the output lacks geometric consistency (third column). E.g., the clock changes shape, and pushing it deeper into the scene fails to reduce its size appropriately; fine details on the can are lost. In contrast, Generative Blocks World infers 3D primitives (fourth column) that can be explicitly manipulated (fifth column), producing a plausible image that respects object geometry, scale, positioning, and texture (last column). We also compare with proprietary models in supplement. \textbf{Second row} Drag Diffusion requires many arrows to place the objects. Notice how objects still do not move precisely where we want them, and there are shape and color mismatches on the rendered watermelon and potato. Our result respects both texture and geometry.
    }
\label{fig:edit_1}
\vspace{-10pt}
\end{figure}

\textbf{Adding Spatial Conditions.} We build upon the state-of-the-art Flux, a rectified flow model~\cite{ esser2024scaling, flux2024}. Older ControlNet implementations~\cite{zhang2023adding} train an auxiliary encoder that adds information to decoder layers of a base frozen U-Net. Newer implementations, including models supplied by the Black Forest Labs developers, concatenate the latent $\mathbf{x_t}$ and condition (e.g., depth map) $\mathbf{c}$ as an input to the network, yielding tighter control. \texttt{FLUX.1 Depth [dev]} re-trains the RF model with the added conditioning; \texttt{FLUX.1 Depth [dev] LoRA} trains LoRA layers on top of a frozen base RF model. Both options give tight control and work well with our primitives, though LoRA exposes an added parameter $lora_{weight} \in [0,1]$ tuning how tightly the depth map should influence synthesis. This is helpful when the primitive abstraction is too coarse relative to the geometric complexity of the desired scene (see Fig.~\ref{fig:LoraAblation}). 

\textbf{Role of Hint and Mask.}
A core contribution of this work is an algorithm to generate a ``hint'' image to guide the image generation process, as well as a confidence mask (see Sec~\ref{sec:texture}). The hint and mask influence the generation within timesteps \( t_{\text{end}} \leq t \leq t_{\text{start}} \), which are hyperparameters. The mask \( \mathbf{m} \in [0,1] \) specifies regions where the hint should guide the output. The hint is encoded into latents \( \mathbf{x}_{\text{hint}} \) via the VAE. During denoising, the latents are updated as $\mathbf{x}_t = (1 - \mathbf{m}) \cdot \mathbf{x}_{\text{hint},t} + \mathbf{m} \cdot \mathbf{x}_t$,
where \( \mathbf{x}_{\text{hint},t} \) is the noised hint latent at timestep \( t \): $\mathbf{x}_{\text{hint},t} = \text{SchedulerScaleNoise}(\mathbf{x}_{\text{hint}}, t, \boldsymbol{\epsilon}).$ Thus, the hint image is \emph{noised} to match the current timestep’s noise level before incorporation, ensuring consistency with the denoising process. Outside \( [t_{\text{end}}, t_{\text{start}}] \), the hint and mask are ignored.

\subsection{Texture Hint Generation for Camera and Object Edits}
\label{sec:texture}

\begin{figure}[t!]
\centering
\includegraphics[width=\linewidth]{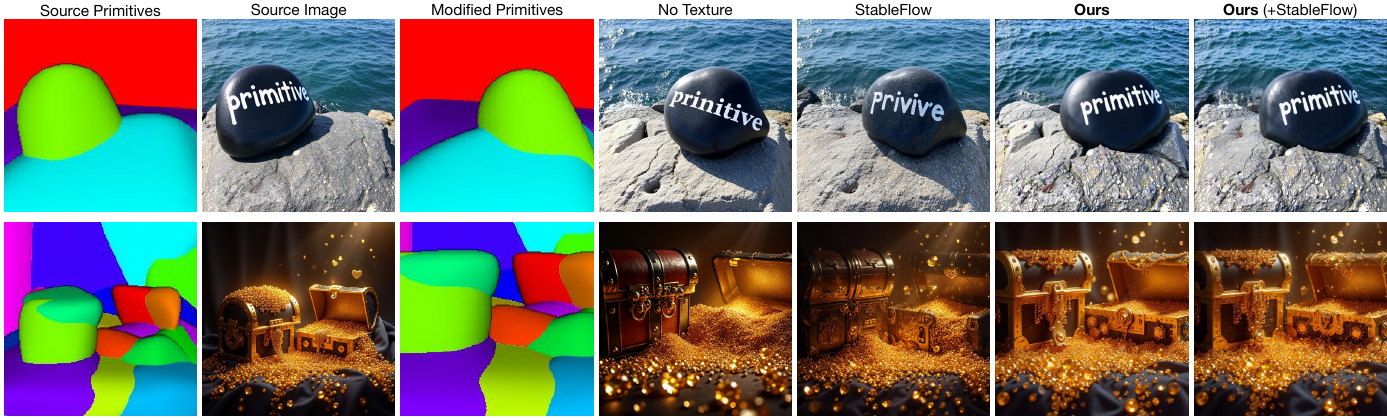}
\vspace{-12pt}
\caption{\textbf{Projection-Based Texture Hints Preserve Object Identity After Edits.}
This figure compares our projection-based texture hints against StableFlow~\cite{avrahami2024stableflow}, which uses vital-layer key-value injection. \textbf{First two columns:} input primitives and image. \textbf{Third:} edited primitives. \textbf{Fourth:} synthesis from original depth, revealing consistent geometry but altered texture. \textbf{Fifth:} StableFlow’s approach often changes texture or object identity. \textbf{Sixth:} our projection-based hints maintain texture fidelity despite edits. \textbf{Seventh:} combining both approaches sometimes improves fine detail recovery (e.g., the treasure chest).
  } 
\label{fig:stableFlowComp}
\vspace{-10pt}
\end{figure}

A number of methods have been proposed to preserve texture/object identity upon editing an image. A common and simple technique is to copy the keys and values from a style image into the newly generated image (dubbed ``style preserving edits''). For older U-Net-based systems, this is done in the bottleneck layers~\cite{LooseC}. For newer DiTs, this is done at selected ``vital'' layers~\cite{avrahami2024stableflow}. In our testing, key-value copying methods are insufficient for camera/primitive moves (see Fig.~\ref{fig:stableFlowComp}). Further, because of our primitives, we have a geometric representation of the scene. Here we demonstrate a routine to obtain a source ``hint'' image $\mathbf{x}_{hint}$ as well as a confidence mask $\mathbf{m}$ that can be incorporated in the diffusion process. The hint image is a rough approximation of what the synthesized image should look like using known spatial correspondences between primitives in the first view and the second. The confidence mask indicates where we can and cannot trust the hint, commonly occurring near depth discontinuities. We rely on the diffusion machinery to essentially clean up the hint, filling gaps and refining blurry projected textures so it looks like a real image. The result of our process is an image that respects the text prompt, source texture, and newly edited primitives/camera.

\textbf{Creating point cloud correspondences} We develop a method that accepts point clouds at the ray-primitive intersection points, a \textit{convex\_map} integer array indicating which primitive was hit at each pixel, a list of per-primitive transforms (such as scale, rotate, translate), and a hyperparameter $max\_distance$ for discarding correspondences. This procedure also robustly handles camera moves because the input point clouds are representations of the same scene in world space.

\textbf{Creating a texture hint} Given a correspondence map of each 3D point in the new view relative to the original view, we can apply this correspondence to generate a hint image that essentially projects pixels in the old view onto the new view. This is the $\mathbf{x_{hint}}$ supplied to the image generation model, taking into account both camera moves and primitive edits like rotation, translation, and scaling. The point cloud correspondence ensures that if a primitive moves, its texture moves with it. In practice, this hint is essential for good texture preservation (see Fig.~\ref{fig:stableFlowComp}). Correspondence and hint generation take about 1-2 seconds per image; 30 denoising steps of FLUX at 512 resolution take about 3 seconds on an H100 GPU. 

\begin{figure*}[t!]
\vspace{-5pt}
  \centering
  \includegraphics[width=\linewidth]{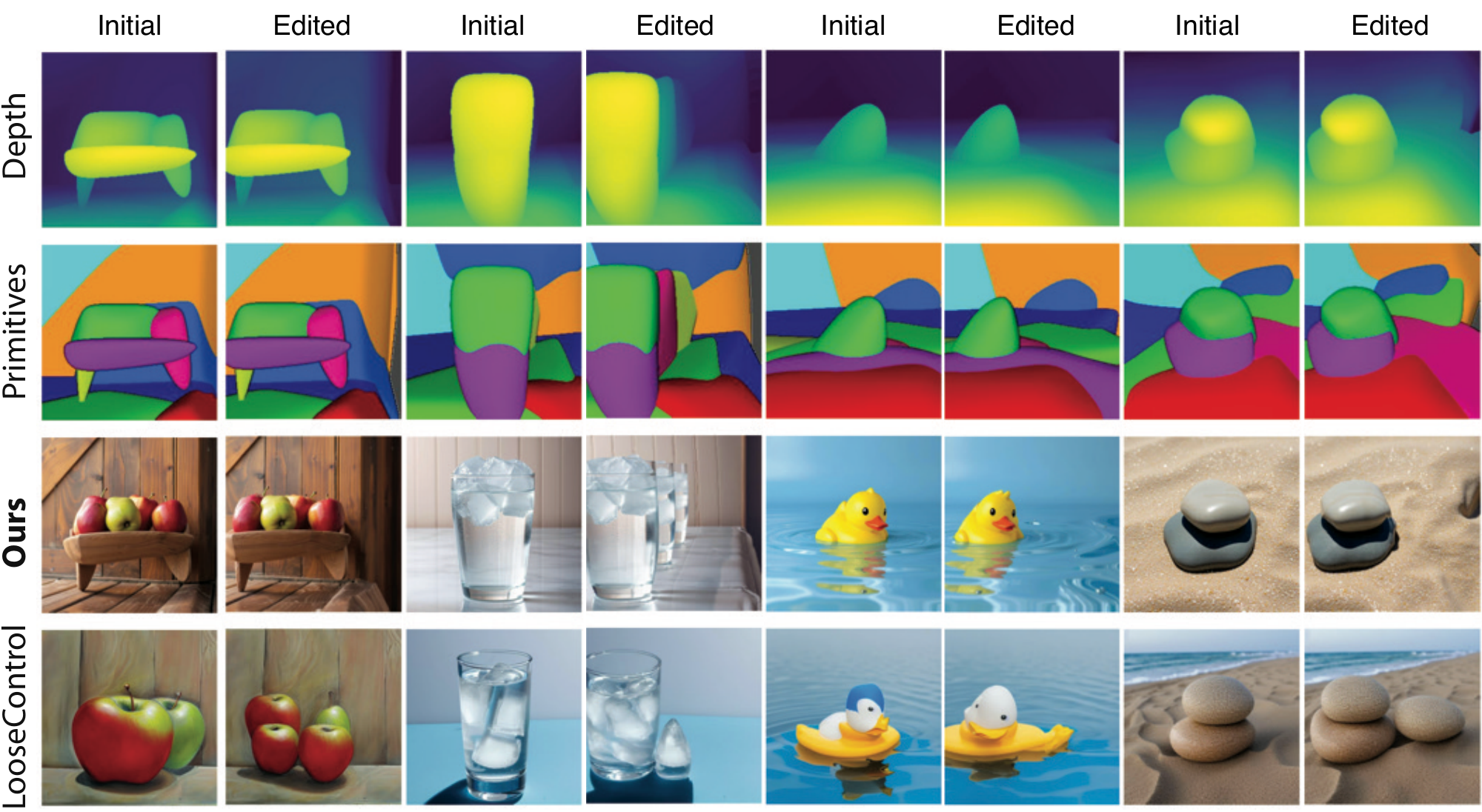}
  \\\vspace{-2pt}
  \caption{\textbf{Comparison with LooseControl}~\cite{LooseC}. Existing work struggles with camera moves. Four scenes ({\bf left} side of each pair), synthesized from
    the depth maps shown.  In each case, the camera is moved to the right ({\bf right} side of each pair), and the image is resynthesized.
    Note how, for LooseControl, the number of apples changes (first pair); the level of water in the glass changes and there is an extra ice cube (second pair);
    the duck changes (third pair); and an extra rock appears (fourth pair). In each case, our method shows the same scene from a different view, because the texture hint image is derived from the underlying geometry, and strongly constrains any change.
    }
  \label{fig:cammovegood}
  \vspace{-5pt}
  \end{figure*}

\subsection{Evaluation}
\label{sec:evals}

We seek error metrics to establish (1) geometric consistency between the primitives requested vs. the image that was synthesized and (2) texture consistency between the source and edited image. For (1) we compute the AbsRel between the depth map supplied to the depth-to-image model (obtained by rendering the primitives) and the estimated depth of the synthesized image (we use the hypersim metric depth module from ~\cite{depth_anything_v2} to get linear depth). Consistent with standard practice in depth estimation, we use least squares to fit scale and shift parameters onto the depth from RGB (letting the primitive depth supplied to the DM be GT). 

To evaluate texture consistency, we apply ideas from the novel view synthesis literature and our existing point cloud correspondence pipeline. Given the source RGB image and the synthesized RGB image (conditioned on the texture hint), we warp the second image back into the first image's frame using our point cloud correspondence algorithm. If we were to synthesize an image in the first render's viewpoint using the second render, this is the texture hint we would use. In error metric calculation, the first RGB image is considered ground truth, the warped RGB image from the edited synthesized image is the prediction, and the confidence mask filters out pixels that are not visible in view 1, given view 2. This evaluation procedure falls in the category of cycle consistency/photometric losses that estimate reprojection error~\cite{jin2024evaluate_geometry,jeong2024nvsadapter,li2025poi,chen2025nvs_depth_priors}.

\begin{table}[t!]
\caption{Comparison of image reconstruction and generation metrics between our method and LooseControl. $\textbf{AbsRel}_{\text{src}}$ and $\textbf{AbsRel}_{\text{dst}}$ are absolute relative errors evaluating how well the generated images adhere to the requested primitive geometry (source and modified, respectively). PSNR and SSIM are evaluated by reprojecting the second synthesized image back to the original camera viewpoint (see Sec~\ref{sec:evals}) and measuring texture consistency with the source. Observe how our procedure simultaneously offers tight geometric adherence to the primitives while preserving the source texture. Results obtained by averaging 48 test images with random camera moves. Because ~\cite{LooseC} does not offer primitive extraction code, we supply our own primitives to both methods for evaluation. We use $K=10$ parts for this evaluation.}
\label{tab:metrics_comparison_revised}
\vspace{-5pt}
\centering
\small
\begin{tabular}{l| cc cc}
\toprule
\textbf{Method} & $\textbf{AbsRel}_{\text{src}}$ $\downarrow$ & $\textbf{AbsRel}_{\text{dst}}$ $\downarrow$ & \textbf{PSNR} $\uparrow$ & \textbf{SSIM} $\uparrow$  \\
\midrule
\textbf{Ours}     & \textbf{0.072} & \textbf{0.076} & \textbf{18.7} & \textbf{0.874}  \\
LooseControl~\cite{LooseC}    & 0.143  & 0.146  & 6.65 & 0.670  \\
\bottomrule
\end{tabular}%
\end{table}

\begin{figure*}[t!]
\vspace{-2pt}
  \centering
  \begin{subfigure}{\linewidth}
    \centering
    \includegraphics[width=\linewidth]{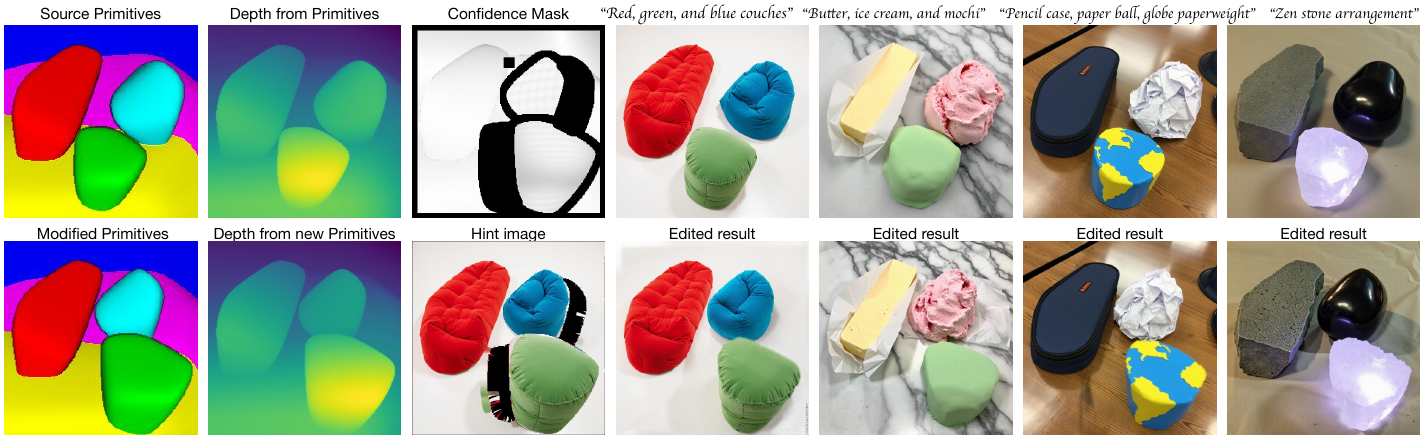}
    \caption{$K=6$ parts.}
    \label{fig:scale_1a}
  \end{subfigure}

  \begin{subfigure}{\linewidth}
    \centering
    \includegraphics[width=\linewidth]{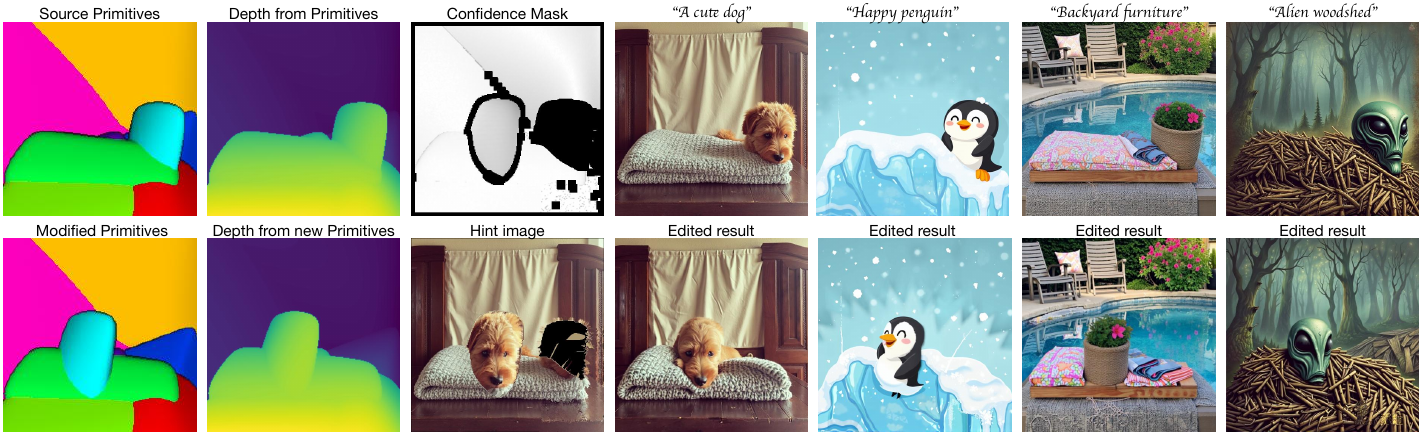}
    \caption{$K=8$ parts.}
    \label{fig:scale_1b}
  \end{subfigure}
  \vspace{-15pt}
  \caption{\textbf{Applying same primitive edit for different text prompts at coarse scale ($K \in \{6,8\}$ parts)}. First row in each subplot contains source primitives and depth (first two columns); the confidence mask for hint generation, followed by four source RGB images. Second row shows the modified primitives and depth, followed by the hint image $x_{hint}$, followed by the four corresponding edited images. At coarse scales, moving a primitive can move a lot of texture at once. Observe how our hint generation procedure automatically yields confidence masks and hints, assigning low confidence to boundaries of primitives that moved (e.g., the dog's hair) and reveals holes when moving objects. The image model cleans up the low-confidence regions and even handles blurry/aliased texture in the hint when $t_{end} > 0$, meaning that the hint is not used for some denoising steps.}
  \label{fig:scale_1}
  \vspace{-10pt}
\end{figure*}

\section{Results}
Fig.~\ref{fig:edit_1} shows how users can manipulate depth map inputs
to depth-to-image synthesizers; Fig~\ref{fig:cammovegood} shows camera
moves. We have precise control over synthesized geometry while respecting
texture. The evaluation in Table~\ref{tab:metrics_comparison_revised},
demonstrates we hit both goals conclusively. Existing texture
preservation based on key-value transfer 
do not preserve details very well, only high-level semantics and style. We ablate the advantage of our texture preservation approach in
Fig.~\ref{fig:stableFlowComp}. When there are few primitives, moving
one primitive affects a big part of the scene; when there are a lot of
primitives, we can make fine-scale edits. We show several such
examples in Figs.~\ref{fig:scale_1}, ~\ref{fig:scale_2} in supplement.

\begin{figure}[t!]
  \centering
  \begin{minipage}[t]{0.50\linewidth}
    \includegraphics[width=\linewidth]{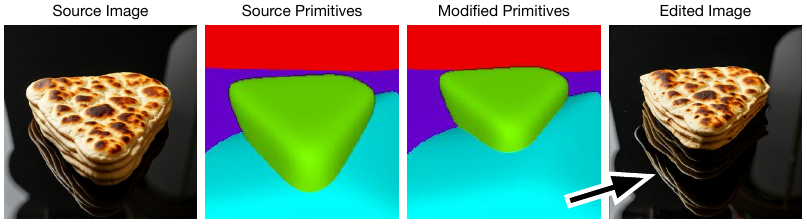}\\[0.5ex]
    \includegraphics[width=\linewidth]{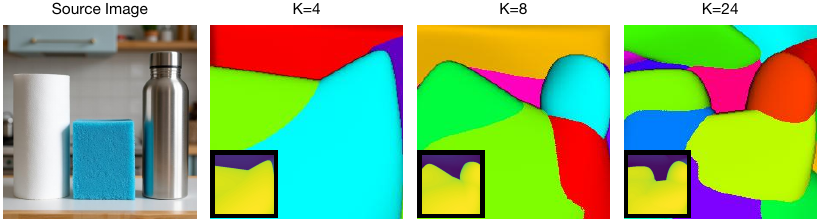}\\[0.5ex]
    \includegraphics[width=\linewidth]{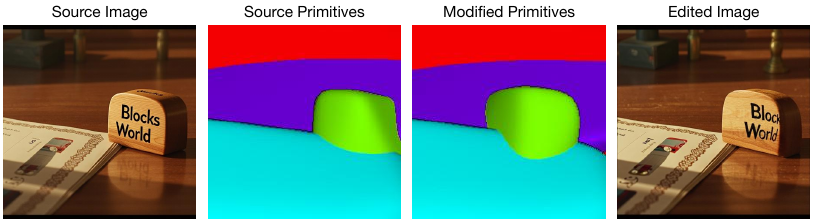}
  \end{minipage}%
  \hfill
  \begin{minipage}[t]{0.45\linewidth}
  \vspace{-58pt}
\caption{
  \textbf{Failure cases.}
  \textbf{Top: Illumination misalignments.} Our pixel-space texture hints fail to model lighting (e.g., reflections, shadows) outside primitive boundaries. Consequently, moving an object like the bread stack does not update its static reflection.
  \textbf{Middle: Poor decomposition.} In cluttered scenes or near image edges, sparse depth can cause primitive fitting to fail, incorrectly merging adjacent objects (bottle and paper towel) and resulting in poor control.
  \textbf{Bottom: Rotation artifacts.} Large object rotations (50 degrees) disrupt geometry and texture consistency, causing distortions or hallucinated content (warped text), likely due to a distribution shift in the texture hints.
}
    \label{fig:failures}
  \end{minipage}
  \vspace{-20pt}
\end{figure}

\section{Discussion}

3D primitives offer precise geometric control over image generation
model outputs, and preserve high-level textures more effectively than
key-value transfer methods. Our method works because
primitive decompositions offer several useful properties:
they are selectable; they are object-linked; they are compact;
they allow edits at coarse and fine grain; and they are accurate enough to yield depth maps that support
high-quality texture projection. Our pipeline is designed to allow
users to choose between coarse and fine control by adjusting the number
of primitives to suit the editing task and scene context. 

Our methods have difficulty with some non-convex shapes (e.g. underside of
a chair or handle of a coffee mug); additional segmentation and
masking, more primitives, or more types of primitive might help.
Depth-of-field blurring/bokeh may not be resolved or sharpened when bringing out-of-focus objects into
focus. Significant object rotations may also fail (see
Fig.~\ref{fig:failures}). In an interactive workflow, manually
expanding the confidence mask to include problematic regions e.g.,
unwanted reflections that don't move with a primitive, can fix some
issues. Future work that applies our point correspondences within the
network layers themselves (e.g., in vital layers) may yield more
robust solutions. Our method does not account for view-dependent
lighting effects and does not enforce temporal consistency across
frames for video synthesis. 

Our work highlights the delicate links between the text prompt, hint
image, initial noise tensor, and depth map. Current inverters do not support our editing model, apparently because
edited images should start from the same noise tensor and prompt as the source image to achieve good
results. Certain edits that are at odds with the text prompt are likely
to cause problems (e.g., if the prompt mentions an object is on the
right, but a user manipulates the primitives to move the object to the
left). Changing the text prompt could work in some circumstances
(Fig.~\ref{fig:fail_caption}).

\subsubsection*{Acknowledgement}
This material is based upon work supported by the National Science Foundation under Grant No. 2106825. This research used both the DeltaAI advanced computing and data resource, which is supported by the National Science Foundation (award OAC 2320345) and the State of Illinois, and the Delta advanced computing and data resource which is supported by the National Science Foundation (award OAC 2005572) and the State of Illinois. Delta and DeltaAI are joint efforts of the University of Illinois Urbana-Champaign and its National Center for Supercomputing Applications.

\bibliography{iclr2026_conference}
\bibliographystyle{iclr2026_conference}
\clearpage
\newpage
\appendix

\section{Appendix}
Here we present additional details and evaluation. Note: we use LLMs as a Latex/Python programming reference and check for related works.  

\subsection{Additional Technical Details}

To lift a depth map \( D \in \mathbb{R}^{H \times W} \) to a 3D point cloud using the pinhole camera model, each pixel \((u, v)\) with depth \( d_{u,v} \) maps to a 3D point \((X, Y, Z)\) as:

\[
X = \frac{(u - c_x) \cdot d_{u,v}}{f_x}, \quad
Y = \frac{(v - c_y) \cdot d_{u,v}}{f_y}, \quad
Z = d_{u,v}
\]

where \( (c_x, c_y) \) is the principal point (typically \( W/2, H/2 \)), and \( (f_x, f_y) \) are the focal lengths along the image axes. DepthAnythingv2 supplies a metric depth module with reasonable camera assumptions. These 3D samples are required to supervise primitive fitting. At test-time, we can directly optimize primitive parameters using the training losses since these 3D samples are available.

\textbf{Primitive fitting details.} We use the standard ResNet-18 encoder (accepting RGBD input) followed by 3 fully-connected layers to predict the parameters of the primitives. We train different networks for different primitive counts $K \in\ \{4,6,8,10,12,24,36,48,60,72\}$, and allow the user to select their desired level of abstraction. Alternatively, the ensembling method of~\cite{vavilala2024improvedconvexdecompositionensembling} can automatically select the appropriate number of primitives. Depending on the primitive count, the training process takes between 40-100 mins on a single A40 GPU, and inference (including generating the initial primitive prediction, refinement, and rendering) can take 1-3 seconds per image. While traditional primitive-fitting to RGB images fits cuboids~\cite{kluger2021cuboids}, we find that polytopes with more faces and without symmetry constraints yield more accurate fits. Thus, we use $F=12$ face polytopes. We do not use a Manhattan World loss or Segmentation loss; the former helped on NYUv2~\cite{Silberman:ECCV12} but not on in-the-wild LAION images and the latter showed an approximately neutral effect in the original paper~\cite{Vavilala_2023_ICCV}.

\begin{table}[htbp!]
\caption{AbsRel depth error metrics for varying numbers of 3D primitives (12-face polytopes). Lower values indicate better depth map approximation quality. While theory would predict AbsRel $\rightarrow 0$ as $K \rightarrow \infty$ (e.g. one primitive per pixel), in practice we run into bias-variance problems fitting more than 60 primitives. Generating primitives is efficient (approx. 1-3 seconds per image on the GPU including finetuning and rendering) so it is feasible for the user to select from a few candidates based on the desired level of abstraction. No other primitive-conditioned image synthesis method offers variable abstraction.}
\centering
\begin{tabular}{cc}
\toprule
\textbf{Number of Parts} ($K$) & \textbf{AbsRel Error}$\downarrow$ \\
\midrule
4  & 0.0376 \\
6  & 0.0330 \\
8  & 0.0295 \\
10 & 0.0282 \\
12 & 0.0265 \\
24 & 0.0223 \\
36 & 0.0203 \\
48 & 0.0202 \\
60 & 0.0194 \\
72 & 0.0195 \\
\bottomrule
\end{tabular}

\label{tab:absrel_vs_parts}
\end{table}

\subsection{Hyperparameter selection}
\label{sec:hyper}
There are a number of hyperparameters associated with our procedure, and we perform a grid search on a held-out validation set to find the best ones. When generating correspondence maps between point clouds, we let \texttt{max\_distance}$=0.005$. In our confidence map, we dilate low-confidence pixels with a score less than $\tau=0.01$ by 9 pixels, which tells the image model to synthesize new texture near primitive boundaries that are often uncertain. We set $(t_{start},t_{end})$ to $(1000,500)$ by default, though $t_{end}$ can be tuned per test image by the user. Applying the hint for all time steps can reduce blending quality near primitive boundaries; not applying the hint for enough time steps could weaken texture consistency. Allowing some time steps to not follow the hint enables desirable super resolution behavior e.g. when bringing a primitive closer to the camera. The supplementary contains detailed algorithms for creating the hint and confidence mask.

\textbf{Inpainting the hint image} After warping the source image to the new view, we find it helpful to inpaint low-confidence regions of the hint $\mathbf{x}_{hint}$ before supplying it to the image model. We considered several possibilities, including \texttt{cv2\_telea} and \texttt{cv2\_ns} from the OpenCV package, as well as simply leaving them as black pixels. We find that Voronoi inpainting, a variation of nearest neighbor inpainting, works well. %
The \texttt{voronoi\_inpainting} function performs image inpainting by filling in regions of low confidence in a hint image using colors from nearby high-confidence pixels, based on a Voronoi diagram approach. 

Given a hint image \( I \) of shape \( [H, W, 3] \) and a confidence mask \( C \) of shape \( [H, W] \) (after resizing if necessary), we identify valid pixels where the confidence satisfies \( C_{i,j} \geq \tau \), with \( \tau \) being the threshold (default 0.01). For each pixel \( (i,j) \) in the image, we assign the color of the nearest valid pixel \( (k,l) \), determined by Euclidean distance, effectively performing nearest-neighbor interpolation. Mathematically, the inpainted image \( I' \) is defined as:
\[
I'_{i,j} = I_{k,l} \quad \text{where} \quad (k,l) = \arg\min_{(m,n) \in V} \sqrt{(i-m)^2 + (j-n)^2},
\]
and \( V = \{(m,n) \mid C_{m,n} \geq \tau\} \) represents the set of high-confidence pixel coordinates. This process leverages a KD-tree for efficient nearest-neighbor searches, ensuring that each pixel adopts the color of the closest reliable pixel, thus preserving local color consistency in the inpainted result. 

For \textbf{FLUX image generation} we begin with the default settings from the diffusers FLUX controlled inpainting pipeline~\footnote{\url{https://huggingface.co/docs/diffusers/en/api/pipelines/control_flux_inpaint}}. We set the \texttt{strength} parameter (controlling starting noise strength) to 1.0 and \texttt{guidance} to 10. We use 30 \texttt{num\_steps} for denoising.
In comparative evaluation, we use the default settings from the authors.

\begin{figure*}[h]
\vspace{-5pt}
\centering
\includegraphics[width=\textwidth]{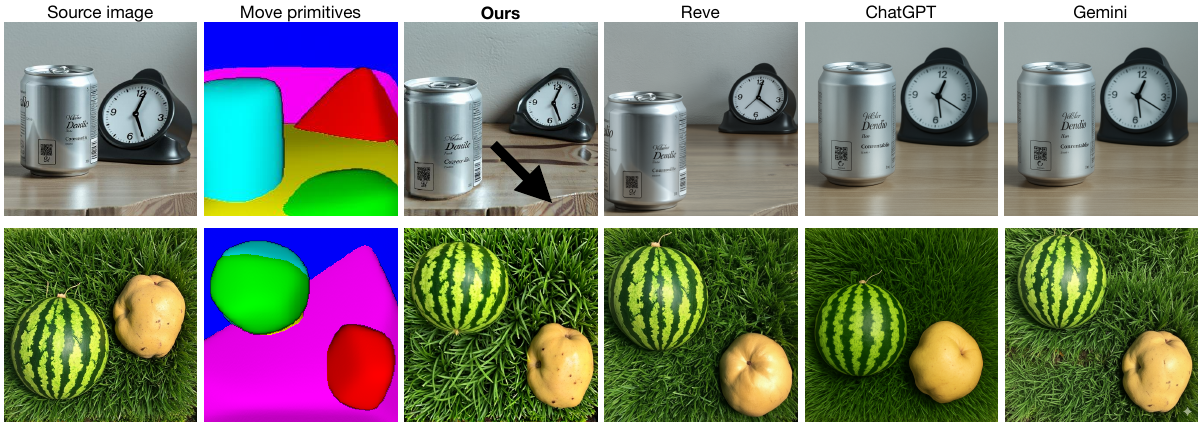}
\caption{\textbf{Evaluation with production systems.} \textbf{First row, Left:} Source image. The next two columns show our primitive edits and the synthesized result. The arrow indicates a texture that our method reproduces faithfully, but others do not. The \textbf{fourth column} shows Reve (https://app.reve.com/), a commercial image generation system. We can prompt their model with 2D boxes to reposition objects, but we must manually estimate their size to take into account 3D perspective effects. With our 3D primitives, maintaining object scale is free. ChatGPT and Gemini do not have interaction mechanisms outside of text prompts and struggle to precisely move objects. Additionally, all 3 production methods added a ``seconds hand'' to the clock that wasn't in the original. Those methods were also unable to generate precise camera moves that we can in this work. The \textbf{second row} shows another example. Our method can precisely move objects while maintaining texture. Reve changed the orientation of the potato. ChatGPT was unable to move the objects where requested (we tried variations of ``move the watermelon to the top left, move the potato to the bottom right"). Gemini succeeded in this example.}
\label{fig:reve}

\vspace{-15pt}
\end{figure*}

\begin{figure*}[h]
  \centering

  \begin{subfigure}{\linewidth}
    \centering
    \includegraphics[width=\linewidth]{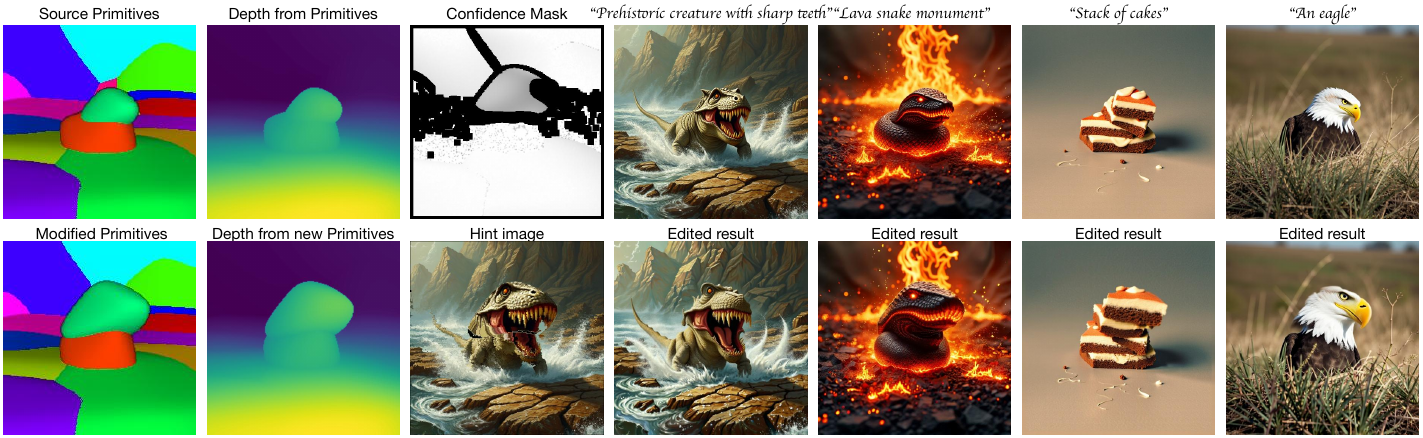}
    \caption{$K=24$ parts.}
    \label{fig:scale_2a}
  \end{subfigure}

  \begin{subfigure}{\linewidth}
    \centering
    \includegraphics[width=\linewidth]{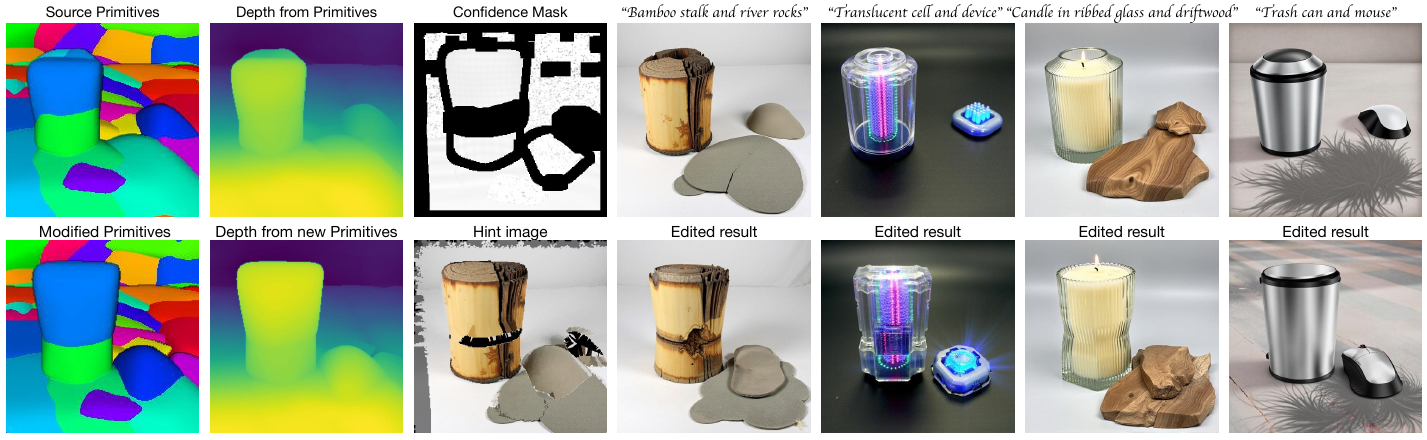}
    \caption{$K=60$ parts.}
    \label{fig:scale_2b}
  \end{subfigure}

  \caption{\textbf{Applying the same primitive edit for different text prompts at fine scale ($K \in \{24,60\}$ parts)}. Observe in the first two rows how all synthesized images respect the enlarged green primitive, while background texture is preserved. In the bottom two rows, we compose \textbf{several edits} using a large number of primitives ($K=60$), enabling fine-scaled edits. We scale up the light blue primitive while scaling down the light green primitive on the left-hand side. We then translate the dark blue primitive on the right-hand side towards the bottom center of the image. We also slightly translate the camera upward. Observe how in the subsequent columns, the edited result respects the geometry specified by the primitives while following the high-level texture of the source image. However, notice how composing four edits challenges our procedure, as the texture preservation isn't as tight. For example, in the final column, a tiled pattern appears on the floor that wasn't in the source.}
  \label{fig:scale_2}
\end{figure*}

\begin{figure*}
  \centering
  \includegraphics[width=\linewidth]{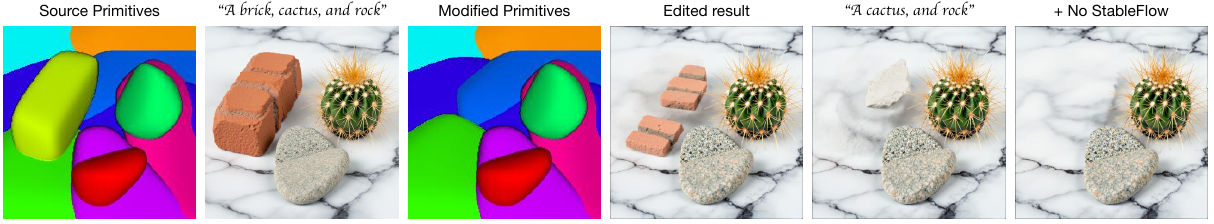}
  \caption{\textbf{Primitive edits can conflict with the text prompt.} Some geometric edits require changing the text prompt, for example, when removing an object. The \textbf{fourth column} mentions brick in the text prompt, but that primitive was removed, resulting in brick pieces in the inpainted region. In the \textbf{fifth column}, we remove the brick from the text prompt, which removes the brick pieces but it still leaves behind a white stone. In the \textbf{final column}, we use our texture hints but without StableFlow, getting a clean surface. The StableFlow key-value sharing approach placed brick and stone textures where we didn't want them. We conclude that our texture hints are critical, but combining them with StableFlow~\cite{avrahami2024stableflow} key-value sharing can help in some cases, hurt in others. 
    }
  \label{fig:fail_caption}
  \vspace{-10pt}
  \end{figure*}

\begin{figure*}[h]
\centering
\includegraphics[width=\linewidth]{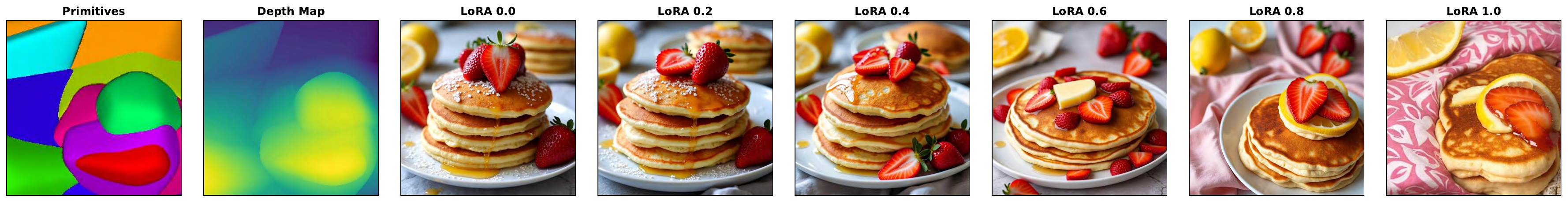}
\includegraphics[width=\linewidth]{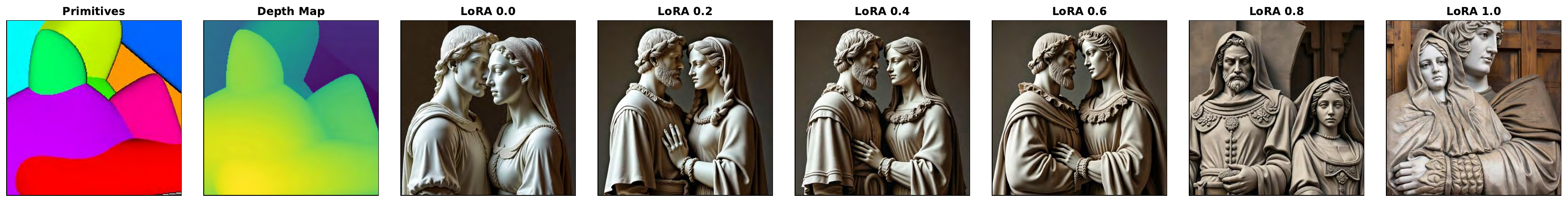}
\includegraphics[width=\linewidth]{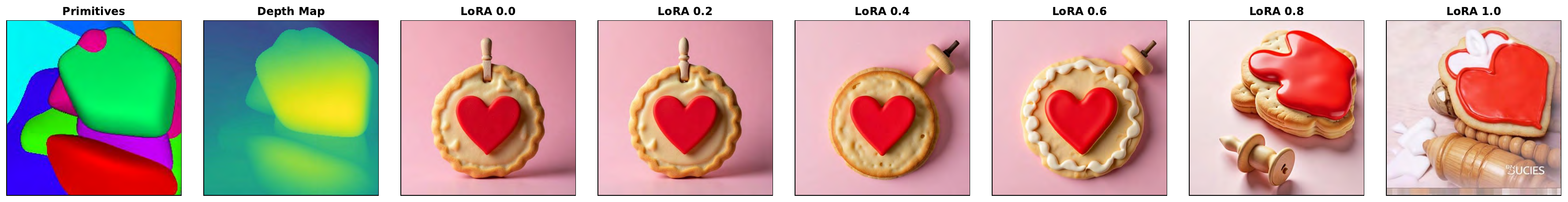}
\includegraphics[width=\linewidth]{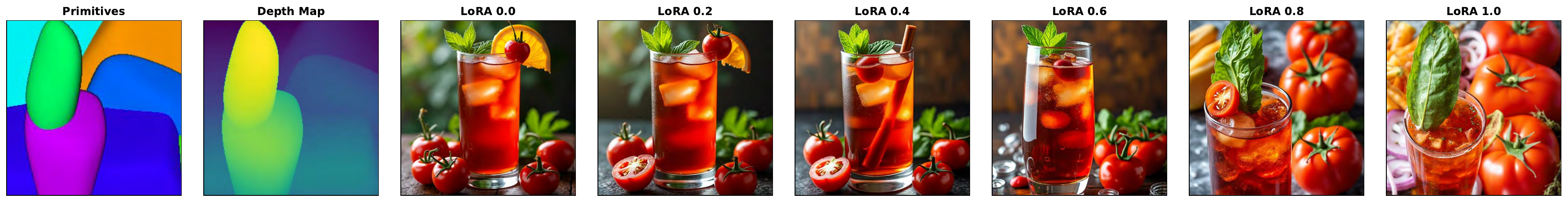}
\caption{Our model is compatible with most depth-image synthesizers. While a pretrained FLUX works out of the box, LoRA weights on top of the base FLUX model are available ( \texttt{FLUX.1 Depth [dev] LoRA}), exposing a new $lora_{weight}$ parameter (scaling the activations of the LoRA layers). This is intriguing in the context of our primitives, because they can either be used to coarsely model scene geometry (e.g. $lora_{weight}$ near 0.8, \textbf{second last column}), leaving details to the image synthesizer, or they can tightly control the result when $lora_{weight}$ is close to 1 (\textbf{final column}).}
\label{fig:LoraAblation}
\end{figure*}

\begin{figure*}[h]
\centering
\includegraphics[width=\linewidth]{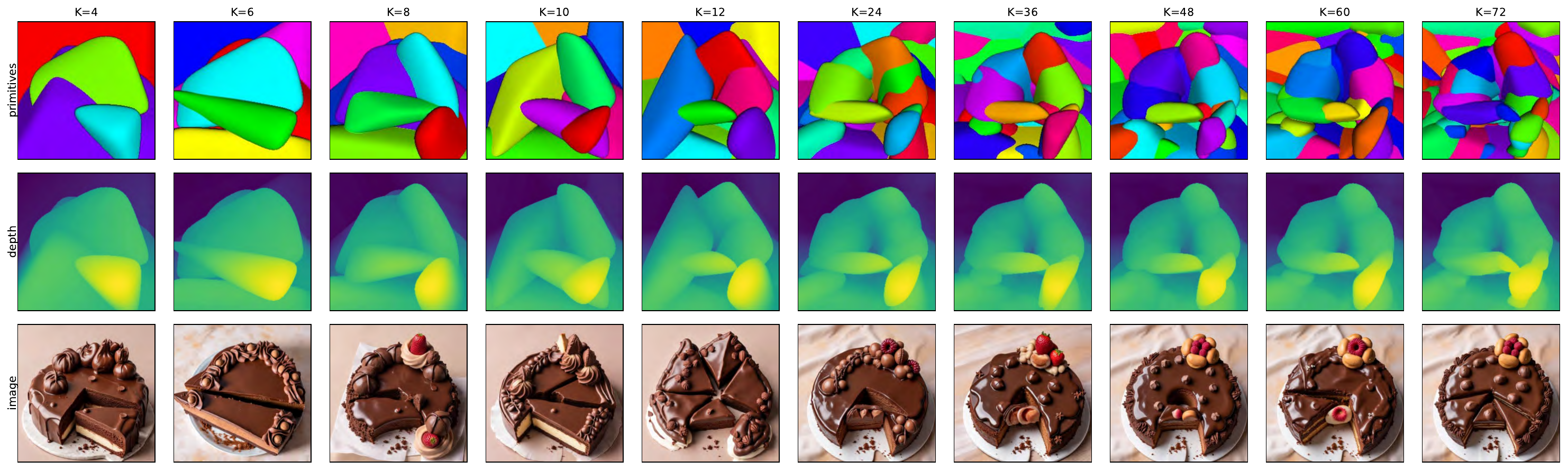}
\includegraphics[width=\linewidth]{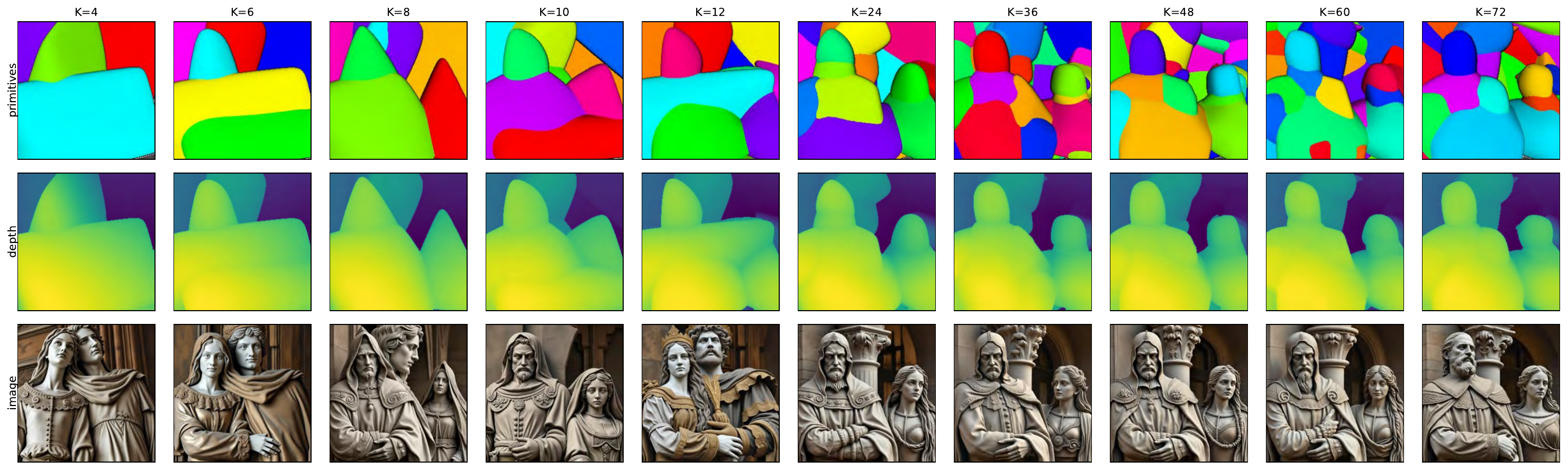}
\caption{Given the same depth map, we extract primitives at variable resolution (from 4-72 parts). We show the depth maps in each second row, and synthesized result in each 3rd row. Observe how no matter the resolution, the FLUX-LoRA model (we use $lora_{weight}=0.8$) gives an image that follows the primitive conditioning. We conclude that a wide array of primitive densities is tolerable to depth-to-image models, enabling meaningful artistic edits.}
\label{fig:PartAbl1}
\end{figure*}

\begin{figure*}[h]
\centering
\includegraphics[width=\linewidth]{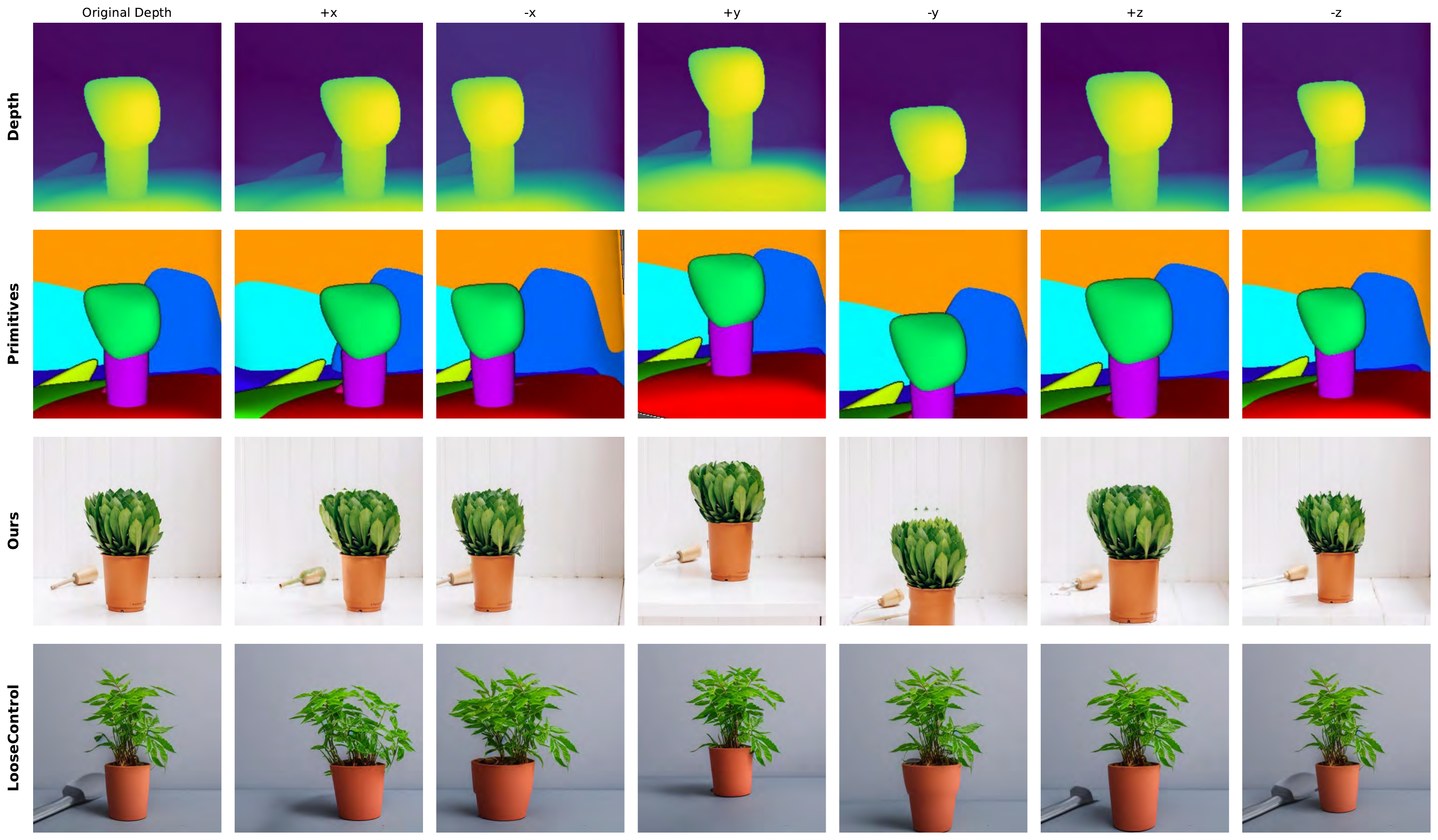}

\caption{Additional move camera evaluations. Our method can simultaneously adhere to source texture and requested primitives.}
\label{fig:eval_LC_4}
\end{figure*}

\begin{figure*}[h]
\centering
\includegraphics[width=\linewidth]{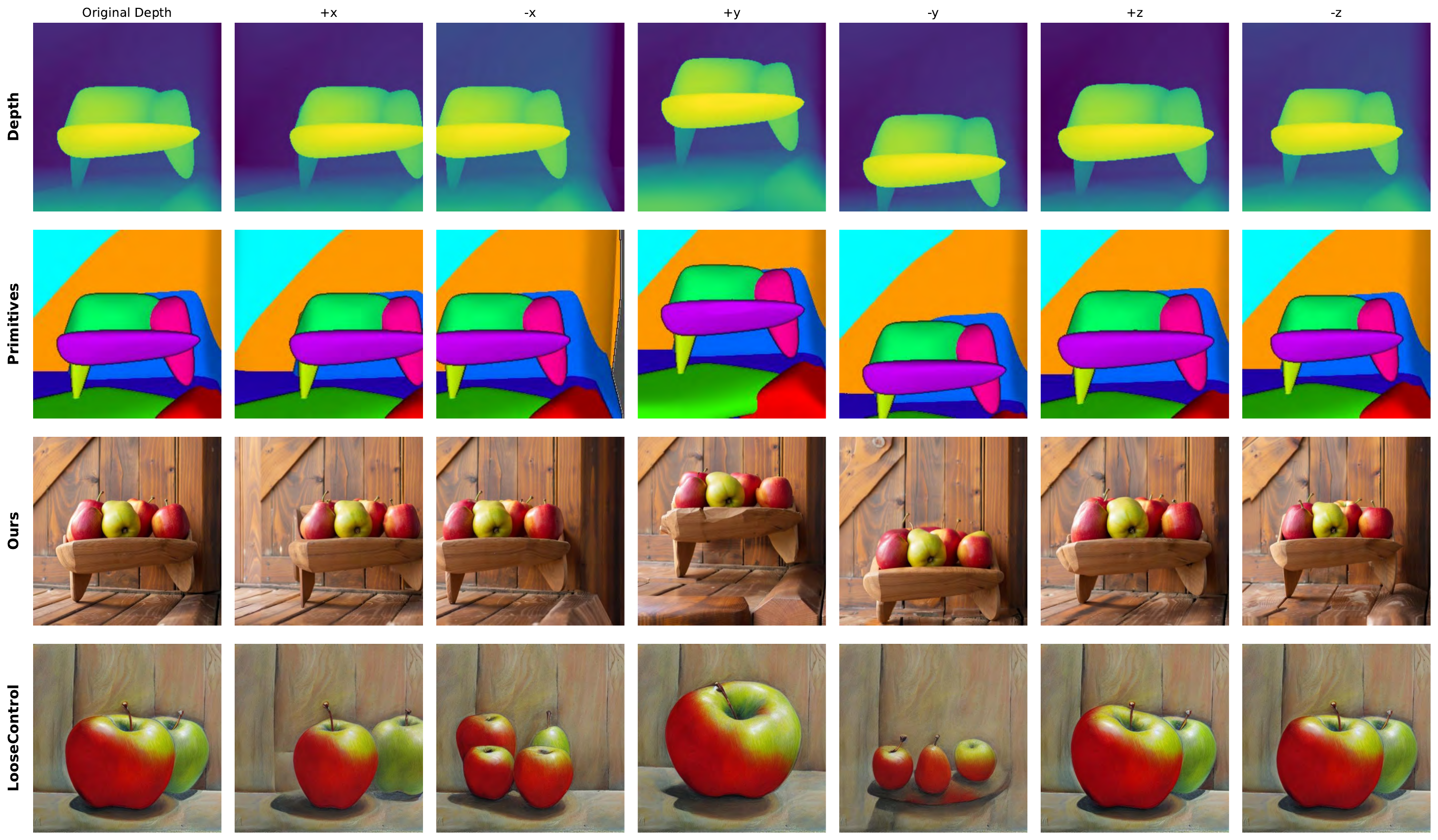}
\includegraphics[width=\linewidth]{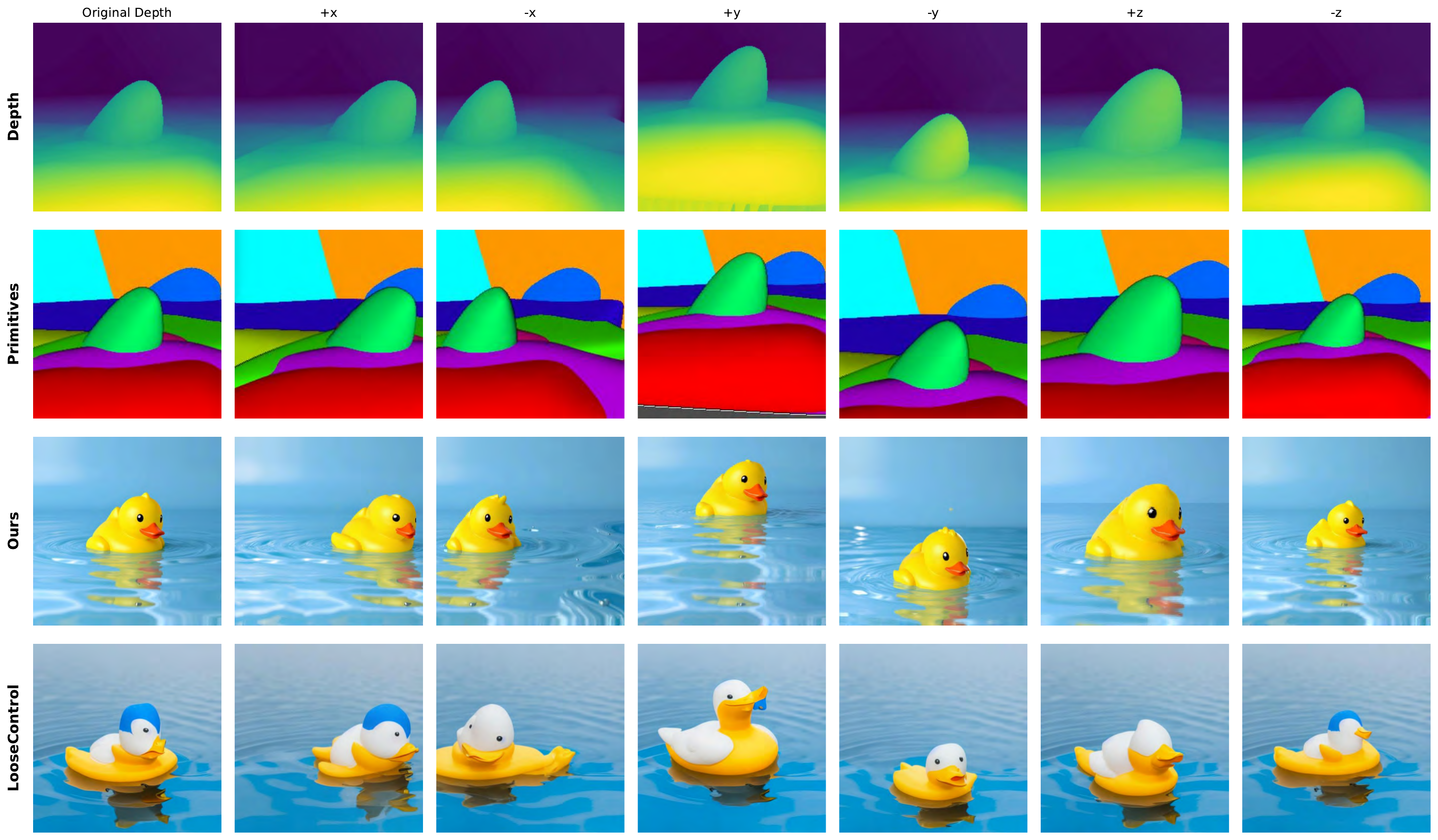}

\caption{Additional move camera evaluations. Generative Blocks World can simultaneously adhere to source texture and requested primitives.}
\label{fig:eval_LC_2}
\end{figure*}

\begin{figure*}[h]
\centering
\includegraphics[width=\linewidth]{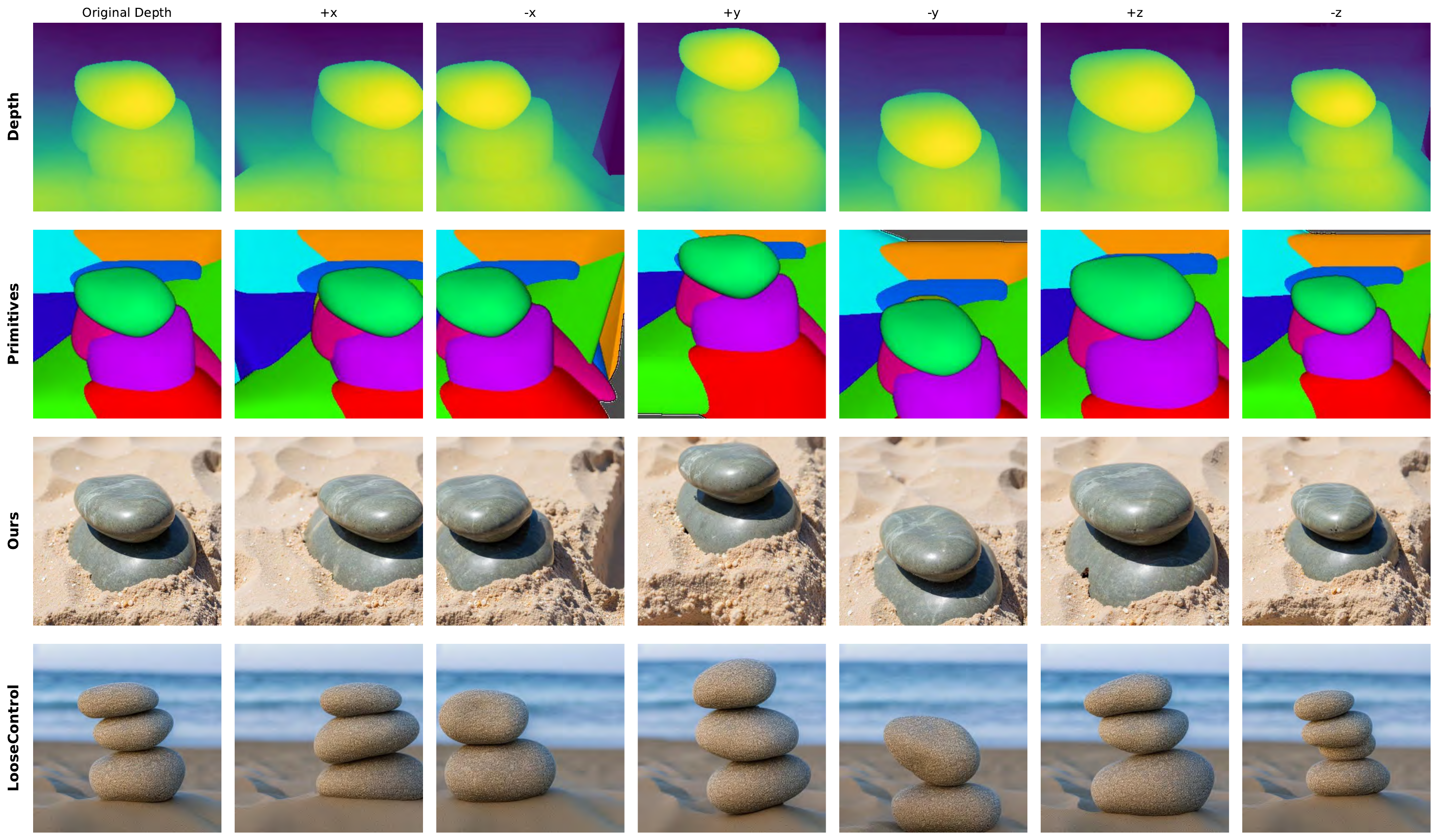}
\includegraphics[width=\linewidth]{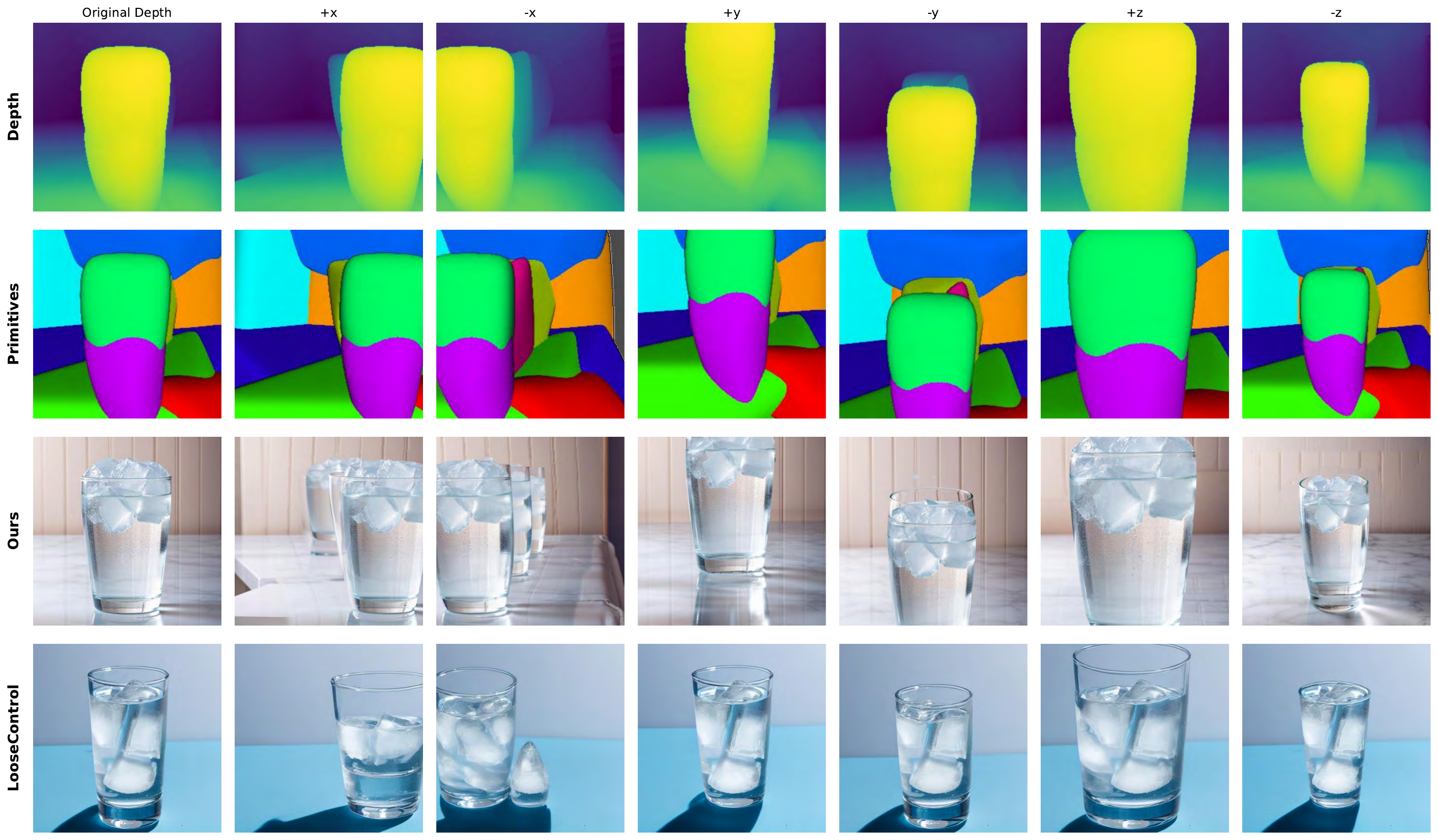}
\caption{Additional move camera evaluations. Our method can simultaneously adhere to source texture and requested primitives.}
\label{fig:eval_LC_3}
\end{figure*}

\begin{figure}[h]
\centering
\includegraphics[width=\linewidth]{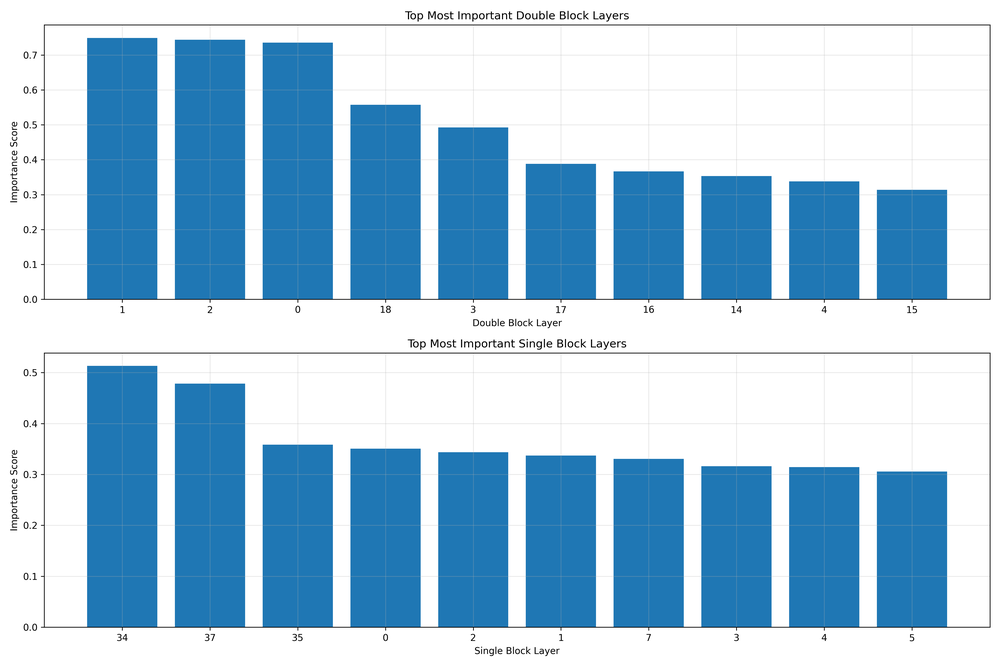}
\includegraphics[width=\linewidth]{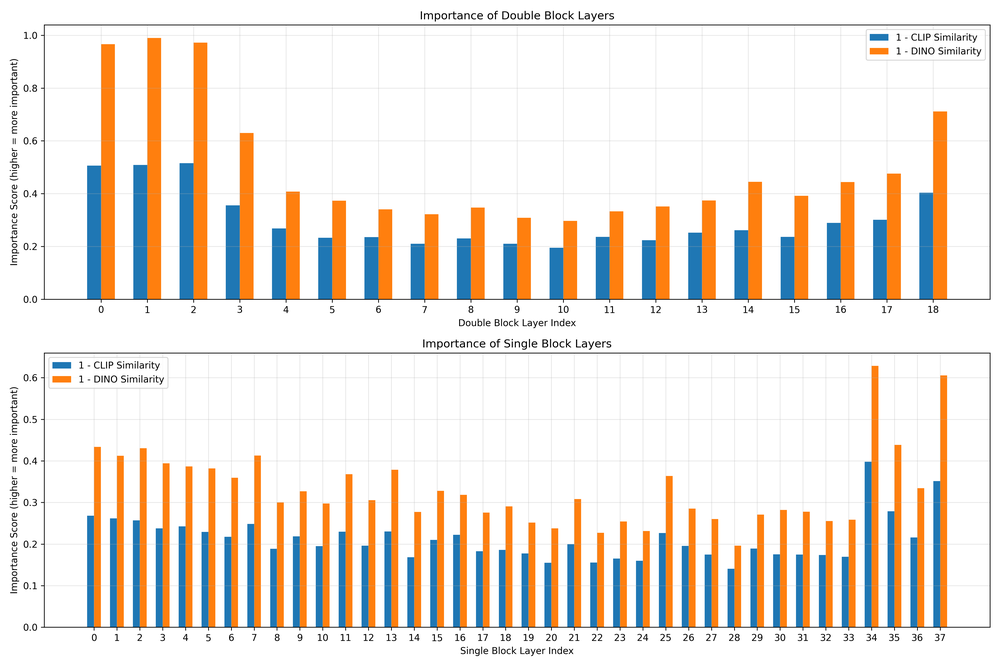}
\caption{We repeat the analysis of StableFlow~\cite{avrahami2024stableflow}, which applies U-Net based key-value transfer of older-generation Diffusion models to newer Diffusion Transformers. Specifically, their work analyzes  \texttt{FLUX.1 [dev]}; given that our work uses depth maps to communicate geometric information to our image generation model, we analyze Vital Layers in  \texttt{FLUX.1 Depth [dev]} and  \texttt{FLUX.1 Depth [dev] LoRA}, finding the top 5 multimodal and single modal layers to be essentially identical. We try using the vital layers we identified for texture transfer, finding this method to be inadequate (see Fig.~\ref{fig:stableFlowComp}).}
\label{fig:stableFlowLayers}
\end{figure}

\begin{algorithm}[t]
\caption{Point Cloud Correspondence Generation}
\label{alg:correspondence}

\KwIn{$\mathcal{P}_1, \mathcal{P}_2$: point clouds; $\mathcal{M}_1, \mathcal{M}_2$: convex maps; $\mathcal{T}$: primitive transforms; $\mathcal{C}$: centers; $d_{\max} = 0.005$: max distance threshold}
\KwOut{$\mathcal{R}$: correspondence map; $\mathcal{W}$: confidence map}

\SetKwFunction{FApplyTransform}{ApplyTransform}
\SetKwProg{Fn}{Function}{:}{}

\BlankLine
\Fn{\FApplyTransform{$\mathbf{p}$, $\mathbf{c}$, $\mathbf{T}$}}{
    $\mathbf{p}' \gets \mathbf{p} - \mathbf{c}$ \tcp*{Center the point}
    
    \If{$\mathbf{T}$ contains translation}{
        $\mathbf{p}' \gets \mathbf{p}' - \mathbf{T}_{\text{trans}}$\;
    }
    
    \If{$\mathbf{T}$ contains rotation angle $\theta$}{
        $c, s \gets \cos(-\theta), \sin(-\theta)$\;
        $x', z' \gets x' \cdot c - z' \cdot s, x' \cdot s + z' \cdot c$ \tcp*{Y-axis rotation}
    }
    
    \If{$\mathbf{T}$ contains scaling factor $scale$}{
        $\mathbf{p}' \gets \mathbf{p}' / scale$\;
    }
    
    \Return $\mathbf{p}' + \mathbf{c}$\;
}

\BlankLine
$\mathcal{R} \gets \mathbf{0}_{H \times W \times 2}$ \tcp*{Initialize correspondence map}
$\mathcal{W} \gets \mathbf{0}_{H \times W}$ \tcp*{Initialize confidence map}

\BlankLine
\For{$p \in \text{unique}(\mathcal{M}_1)$}{
    \If{$p < 0$ \textbf{or} $p \geq |\mathcal{C}|$ \textbf{or} $p \notin \mathcal{M}_1$ \textbf{or} $p \notin \mathcal{M}_2$}{
        \textbf{continue}\;
    }
    
    \BlankLine
    $\mathcal{I}_1 \gets \{(y,x) : \mathcal{M}_1[y,x] = p\}$ \tcp*{Pixel indices for primitive $p$ in map 1}
    $\mathcal{I}_2 \gets \{(y,x) : \mathcal{M}_2[y,x] = p\}$ \tcp*{Pixel indices for primitive $p$ in map 2}
    
    $\mathcal{Q}_1 \gets \{\mathcal{P}_1[y,x] : (y,x) \in \mathcal{I}_1\}$ \tcp*{3D points for primitive $p$}
    
    \BlankLine
    \For{$(y_2, x_2) \in \mathcal{I}_2$}{
        $\mathbf{q} \gets \mathcal{P}_2[y_2, x_2]$ \tcp*{Query point from second cloud}
        
        \If{$p \in \mathcal{T}$}{
            $\mathbf{q} \gets \FApplyTransform(\mathbf{q}, \mathcal{C}[p], \mathcal{T}[p])$ \tcp*{Apply transformation}
        }
        
        \BlankLine
        $\mathbf{d} \gets \|\mathcal{Q}_1 - \mathbf{q}\|_2$ \tcp*{Compute distances to all points}
        $i^* \gets \arg\min_i \mathbf{d}[i]$ \tcp*{Find nearest neighbor}
        $d_{\min} \gets \mathbf{d}[i^*]$\;
        
        \If{$d_{\min} \leq d_{\max}$}{
            $(y_1^*, x_1^*) \gets \mathcal{I}_1[i^*]$ \tcp*{Get corresponding pixel coordinates}
            $\mathcal{R}[y_2, x_2] \gets [x_1^*, y_1^*]$\;
            $\mathcal{W}[y_2, x_2] \gets 1 - \min(d_{\min} / d_{\max}, 1)$ \tcp*{Confidence score}
        }
    }
}

\BlankLine
\Return $\mathcal{R}, \mathcal{W}$\;

\end{algorithm}

\newpage

\begin{algorithm}[t]
\caption{Hint Generation from Correspondence Maps}
\label{alg:hint_generation}

\KwIn{$\mathbf{I}_{\text{src}} \in \mathbb{R}^{C \times H_s \times W_s}$: source image; $\mathcal{R} \in \mathbb{R}^{H_r \times W_r \times 2}$: correspondence map; $\mathcal{W} \in \mathbb{R}^{H_r \times W_r}$: confidence map; $\mathcal{M}_{\text{hit}} \in \{0,1\}^{H_r \times W_r}$: hit mask}
\KwOut{$\mathcal{H} \in \mathbb{R}^{C \times H_s \times W_s}$: generated hint image}

\SetKwFunction{FBilinear}{BilinearSample}
\SetKwProg{Fn}{Function}{:}{}

\BlankLine
\Fn{\FBilinear{$\mathbf{I}$, $y$, $x$}}{
    $C, H, W \gets \text{shape}(\mathbf{I})$\;
    $x \gets \text{clip}(x, 0, W - 1.001)$, $y \gets \text{clip}(y, 0, H - 1.001)$\;
    
    \BlankLine
    $x_0, y_0 \gets \lfloor x \rfloor, \lfloor y \rfloor$ \tcp*{Floor coordinates}
    $x_1, y_1 \gets \min(x_0 + 1, W-1), \min(y_0 + 1, H-1)$\;
    
    $w_x, w_y \gets x - x_0, y - y_0$ \tcp*{Interpolation weights}
    
    \BlankLine
    $\mathbf{v}_{\text{top}} \gets \mathbf{I}[:, y_0, x_0] \cdot (1 - w_x) + \mathbf{I}[:, y_0, x_1] \cdot w_x$\;
    $\mathbf{v}_{\text{bot}} \gets \mathbf{I}[:, y_1, x_0] \cdot (1 - w_x) + \mathbf{I}[:, y_1, x_1] \cdot w_x$\;
    
    \Return $\mathbf{v}_{\text{top}} \cdot (1 - w_y) + \mathbf{v}_{\text{bot}} \cdot w_y$\;
}

\BlankLine
$\lambda_h \gets H_s / H_r$, $\lambda_w \gets W_s / W_r$ \tcp*{Scale factors}
$\mathcal{H} \gets \mathbf{0}_{C \times H_s \times W_s}$ \tcp*{Initialize hint image}

\BlankLine
\For{$y \in [0, H_r)$}{
    \For{$x \in [0, W_r)$}{
        \If{$\mathcal{M}_{\text{hit}}[y, x] = 1$}{
            \textbf{continue} \tcp*{Skip hit pixels}
        }
        
        \BlankLine
        $(x_c, y_c) \gets \mathcal{R}[y, x]$ \tcp*{Get correspondence}
        $w \gets \mathcal{W}[y, x]$ \tcp*{Get confidence}
        
        \If{$w < 0.1$}{
            \textbf{continue} \tcp*{Skip low-confidence correspondences}
        }
        
        \BlankLine
        $y_{\text{src}} \gets y_c \cdot \lambda_h$, $x_{\text{src}} \gets x_c \cdot \lambda_w$ \tcp*{Scale to source resolution}
        
        $y_{\text{start}} \gets \lfloor y \cdot \lambda_h \rfloor$, $y_{\text{end}} \gets \lfloor (y+1) \cdot \lambda_h \rfloor$\;
        $x_{\text{start}} \gets \lfloor x \cdot \lambda_w \rfloor$, $x_{\text{end}} \gets \lfloor (x+1) \cdot \lambda_w \rfloor$\;
        
        \BlankLine
        \For{$y_s \in [y_{\text{start}}, y_{\text{end}})$}{
            \For{$x_s \in [x_{\text{start}}, x_{\text{end}})$}{
                \If{$y_s \notin [0, H_s)$ \textbf{or} $x_s \notin [0, W_s)$}{
                    \textbf{continue} \tcp*{Boundary check}
                }
                
                \BlankLine
                $\alpha_y \gets \frac{y_s - y_{\text{start}}}{\max(y_{\text{end}} - y_{\text{start}}, 1)}$ \tcp*{Normalized offset}
                $\alpha_x \gets \frac{x_s - x_{\text{start}}}{\max(x_{\text{end}} - x_{\text{start}}, 1)}$\;
                
                $y_{\text{sample}} \gets y_{\text{src}} + \alpha_y \cdot \lambda_h$\;
                $x_{\text{sample}} \gets x_{\text{src}} + \alpha_x \cdot \lambda_w$\;
                
                $\mathcal{H}[:, y_s, x_s] \gets \FBilinear(\mathbf{I}_{\text{src}}, y_{\text{sample}}, x_{\text{sample}})$\;
            }
        }
    }
}

\BlankLine
\Return $\mathcal{H}$\;

\end{algorithm}

\clearpage

\subsection{Related work matrix}

\begin{table}[h!]
\vspace{-2pt}
\caption{Related work comparison summary. While all methods can move objects, ours is the only one that uses 3D primitives (at varying density) in a training-free pipeline, while also supporting scene-level camera moves. Previous drag-based works use 2D arrows in pixel space to prompt the edit (combined with spatial masks); in contrast, our approach is prompted via selecting and moving 3D primitives. Loose Control is the closest approach to ours that uses 3D boxes, but it requires training and cannot support detailed, variable-density primitives. Here, Variable LoD refers to support for variable primitive count, to enable both coarse and fine edits. Note that while our image generation process is training-free, our primitives are learned.
}
\vspace{-8pt}
\centering
\resizebox{\columnwidth}{!}{%
\renewcommand{\arraystretch}{1.2}
\setlength{\tabcolsep}{6pt}
\begin{tabular}{lccccc}
\toprule
\textbf{Method} & \textbf{Interaction} & \textbf{Training-free} & \textbf{Move Objects} & \textbf{Variable LoD} & \textbf{Camera Move} \\
\midrule
Diff. Self-Guid.~\cite{epstein2023selfguidance} & 2D guidance & \cmark & \cmark & \xmark & \xmark \\
Diffusion Handles~\cite{10657993} & 3D handles & \cmark & \cmark & \xmark & \xmark \\
Edit. Image Elements~\cite{mu2024editableECCV} & 2D elements & \xmark & \cmark & \xmark & \xmark \\
DragDiffusion~\cite{shi2024dragdiffusion} & 2D points & \xmark & \cmark & \xmark & \xmark \\
GoodDrag~\cite{zhang2025gooddrag} & 2D points & \cmark & \cmark & \xmark & \xmark \\
FlowDrag~\cite{koo2025flowdrag} & 2D points & \cmark & \cmark & \xmark & \xmark \\
DragIn3D~\cite{Guang_2025_CVPR} & 3D points & \xmark & \cmark & \cmark & \xmark \\
\midrule
LooseControl~\cite{LooseC} & 3D boxes & \xmark & \cmark & \xmark & \cmark \\
\textbf{Ours} & 3D primitives & \cmark & \cmark & \cmark & \cmark \\
\bottomrule
\end{tabular}%
}
\vspace{-12pt}
\end{table}

\section{Additional Evaluation}
\setcounter{page}{1}

\begin{table}[h]
\centering
\caption{\textbf{Additional ablations evaluating geometric consistency and texture consistency on the open DragBench dataset}. Similar to Table~\ref{tab:metrics_comparison_revised}, we evaluate novel-view consistency. AbsRel$_{\text{src}}$ and AbsRel$_{\text{dst}}$ measure depth consistency for source and destination views respectively. Best values are shown in \textbf{bold}. $\uparrow$ indicates higher is better; $\downarrow$ indicates lower is better. This experiment uses $K=10$ parts, and on average 78.1\% of pixels were visible in both views. Notice how our texture-hint mechanism is essential for good results; StableFlow~\cite{avrahami2024stableflow} is occasionally helpful. See Fig.~\ref{fig:stableFlowComp} for qualitative examples.}
\label{tab:model_comparison}

\caption*{\footnotesize DragBench dataset: \href{https://huggingface.co/datasets/LuJingyi/DragBench}{https://huggingface.co/datasets/LuJingyi/DragBench}}

\begin{tabular}{lccccc}
\toprule
Method & PSNR $\uparrow$ & SSIM $\uparrow$ & LPIPS $\downarrow$ & AbsRel$_{\text{src}}$ $\downarrow$ & AbsRel$_{\text{dst}}$ $\downarrow$ \\
\midrule
LooseControl~\cite{LooseC} & 7.36 & 0.501 & 0.335 & 0.148 & 0.151 \\
\textbf{Ours} FLUX (No Hint) & 8.42 & 0.465 & 0.373 & \textbf{0.099} & 0.107 \\
\textbf{Ours} FLUX + StableFlow & 9.11 & 0.477 & 0.331 & \textbf{0.099} & 0.107 \\
\textbf{Ours} FLUX + Hint & 14.24 & \textbf{0.658} & \textbf{0.165} & \textbf{0.099} & \textbf{0.102} \\
\textbf{Ours} FLUX + Hint + StableFlow & \textbf{14.46} & 0.654 & 0.169 & \textbf{0.099} & 0.113 \\
\bottomrule
\end{tabular}
\end{table}

\begin{figure}[h]
\centering
\includegraphics[width=\linewidth]{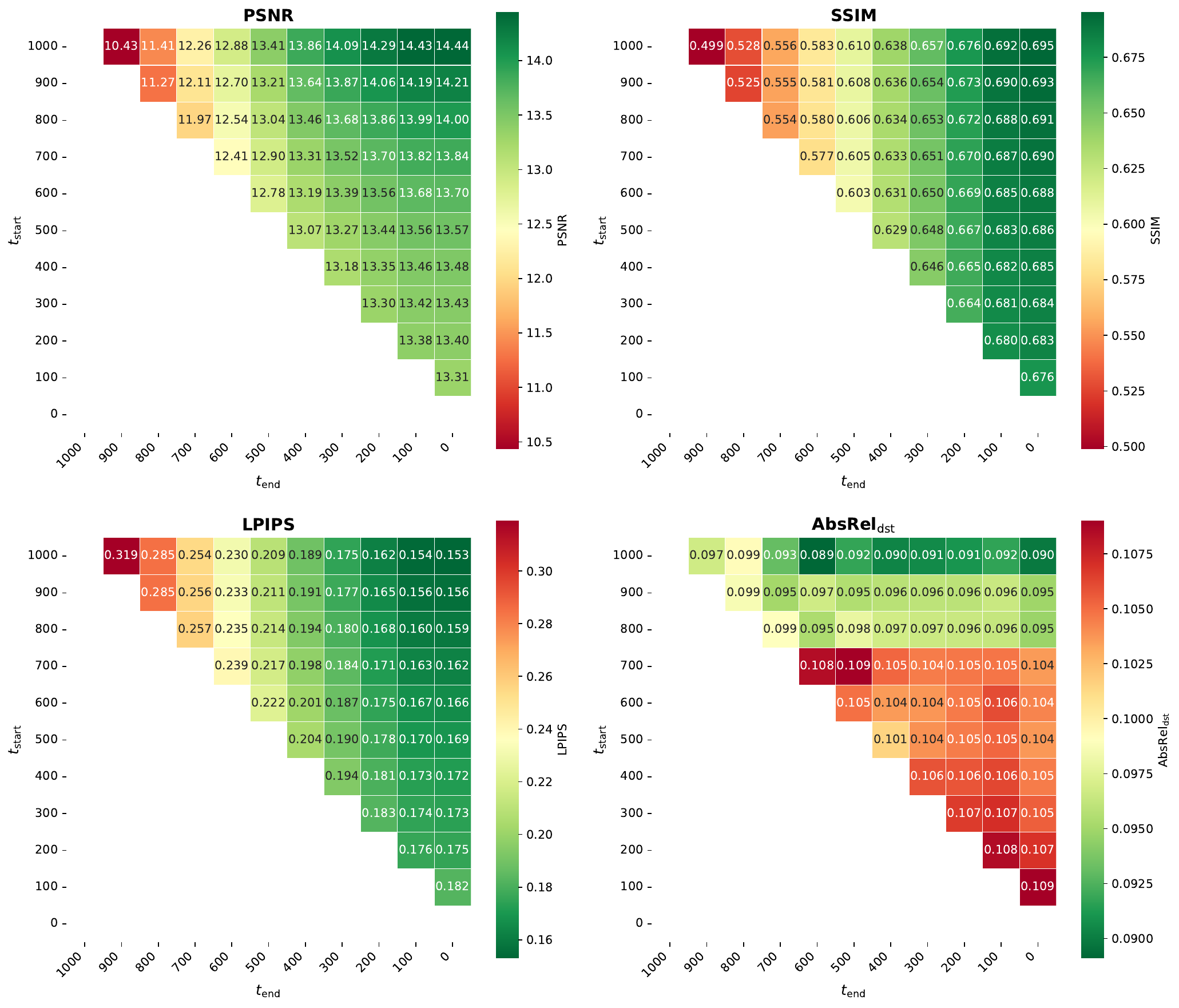}
\caption{\textbf{Grid search over hint-timestep parameters $t_{\text{start}}$ and $t_{\text{end}}$}. The hint image steers the generation only within these timesteps, and only for pixels in the confidence mask. This evaluation suggests that $t_{\text{start}} \approx 1000$ and $t_{\text{end}} \approx 0$ obtain the best overall texture-consistency metrics in regions shared between source and destination views (as expected, since the hint is applied at every step). However, we observed unwanted artifacts when applying the hint throughout the full trajectory, as warped pixels fail to harmonize with inpainted regions. Through qualitative inspection (see Fig.~\ref{fig:hint_qual}), we find $t_{\text{start}}=1000$, $t_{\text{end}}=300$ works well (and is user-tunable at test time); this quantitative evaluation confirms we sacrifice little in texture consistency or depth accuracy.}
\label{fig:tstart_ablation}
\end{figure}

\begin{figure}[h]
\centering
\includegraphics[width=0.7\linewidth]{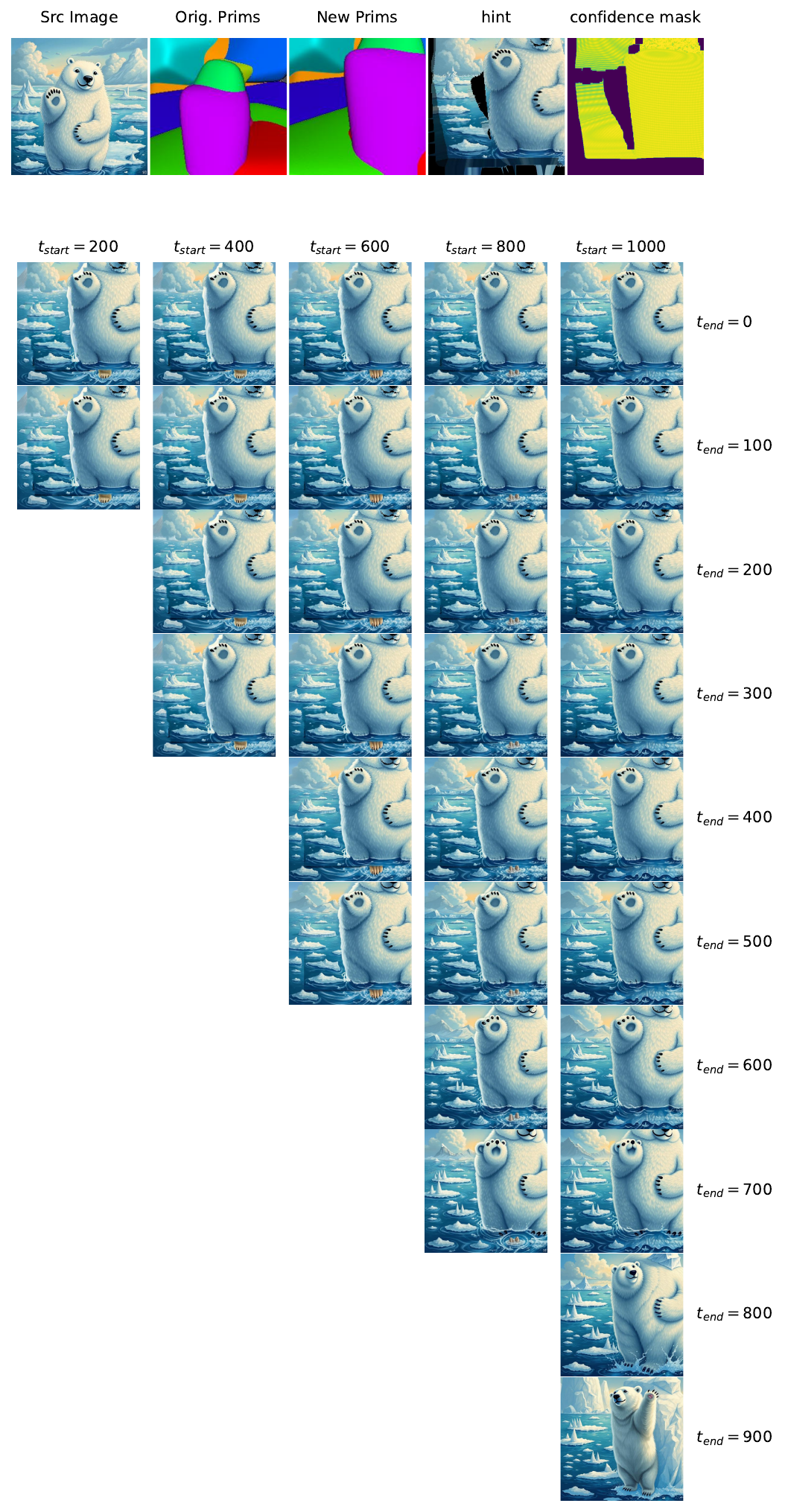}
\caption{\textbf{Qualitative ablation over hint-timestep parameters $t_{\text{start}}$ and $t_{\text{end}}$}. When $t \in [t_{\text{start}}, t_{\text{end}}]$, the hint image influences synthesis. The hint image is essential for steering the model toward texture-consistent solutions, as shown in Table~\ref{tab:model_comparison}. However, it introduces warping and aliasing artifacts that the diffusion process can correct. For example, when $t_{end} < 200$, aliasing appears near the bear’s mouth. If $t_{end}$ is too large ($>700$), the model diverges from the hint (e.g., the bear’s head pose changes). Similarly, if the hint begins too late ($t_{start} < 800$), unwanted content is synthesized (e.g., the orange paw in the ocean). Both quantitative metrics and visual results favor $(t_{\text{start}}, t_{\text{end}}) \approx (1000, 300)$.}
\label{fig:hint_qual}
\end{figure}

\begin{figure}[h]
\centering
\includegraphics[width=\linewidth]{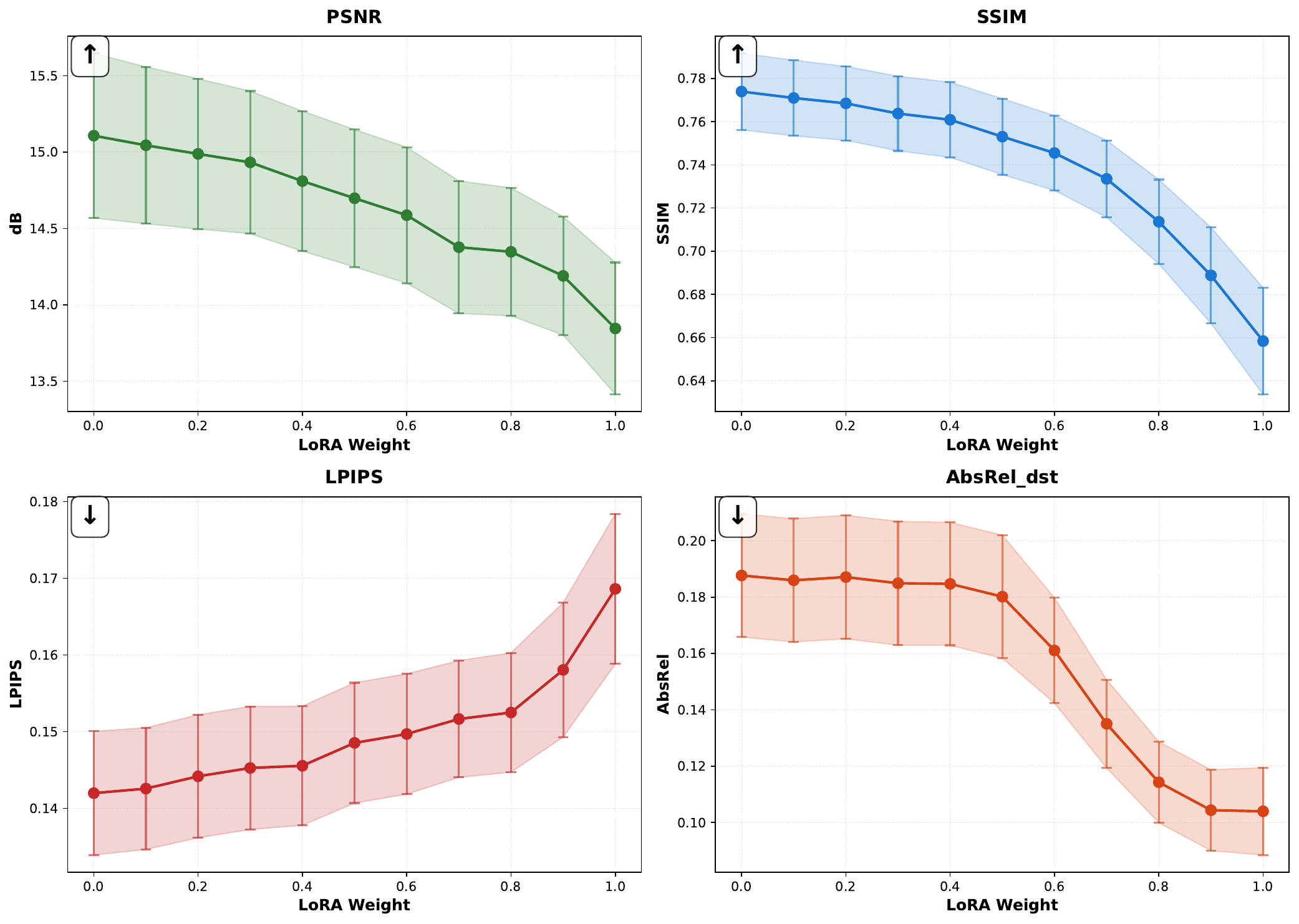}
\caption{\textbf{Ablation of LoRA depth-conditioning weight on texture and geometry.} Error bars show 95\% confidence intervals. As the LoRA weight increases, texture similarity to the hint image steadily degrades across LPIPS, SSIM, and PSNR, revealing the inherent difficulty of achieving strong texture consistency and strong depth consistency simultaneously. Increasing LoRA weight strengthens depth-specific pathways in the diffusion model, improving geometric faithfulness—evidenced by the monotonic reduction in AbsRel up to weight = 0.9—but at the cost of reduced adherence to the hint image. This highlights an intrinsic tradeoff between geometric accuracy and texture preservation, and the LoRA weight provides a simple, interpretable knob to tune this balance. See Fig.~\ref{fig:EditLoraAblation} for qualitative examples.}
\label{fig:lora_ablate_chart}
\end{figure}

\begin{figure*}[h]
\centering

\begin{minipage}{\linewidth}
    \centering
    \includegraphics[width=\linewidth]{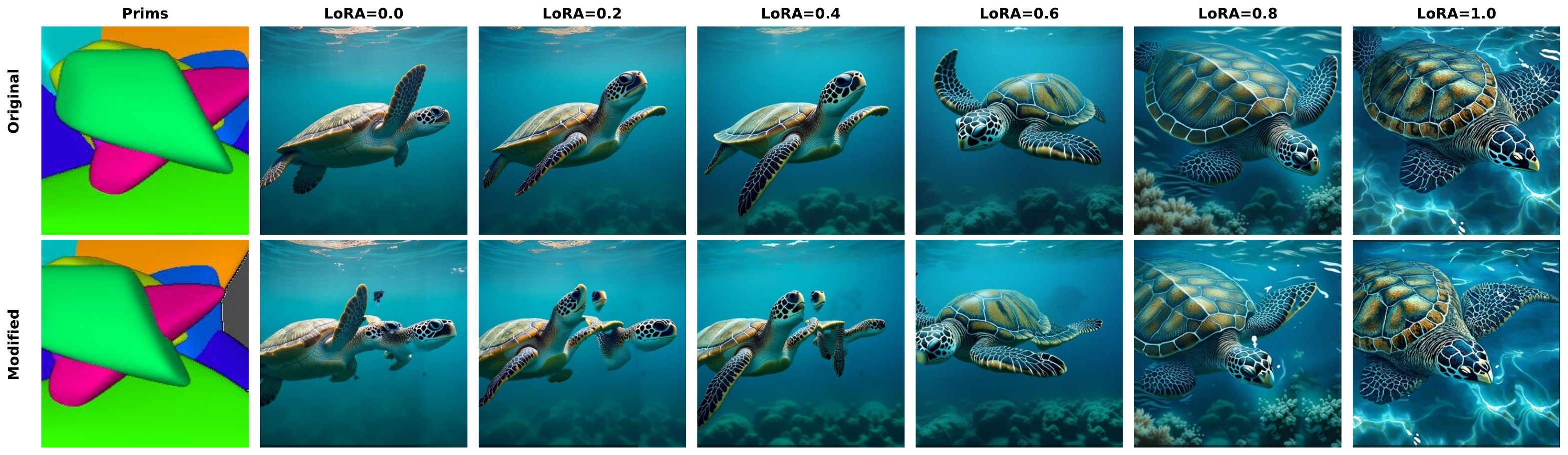}\\[2mm]
    {\fontfamily{pzc}\selectfont ``Turtle is swimming''}
\end{minipage}

\vspace{7mm}

\begin{minipage}{\linewidth}
    \centering
    \includegraphics[width=\linewidth]{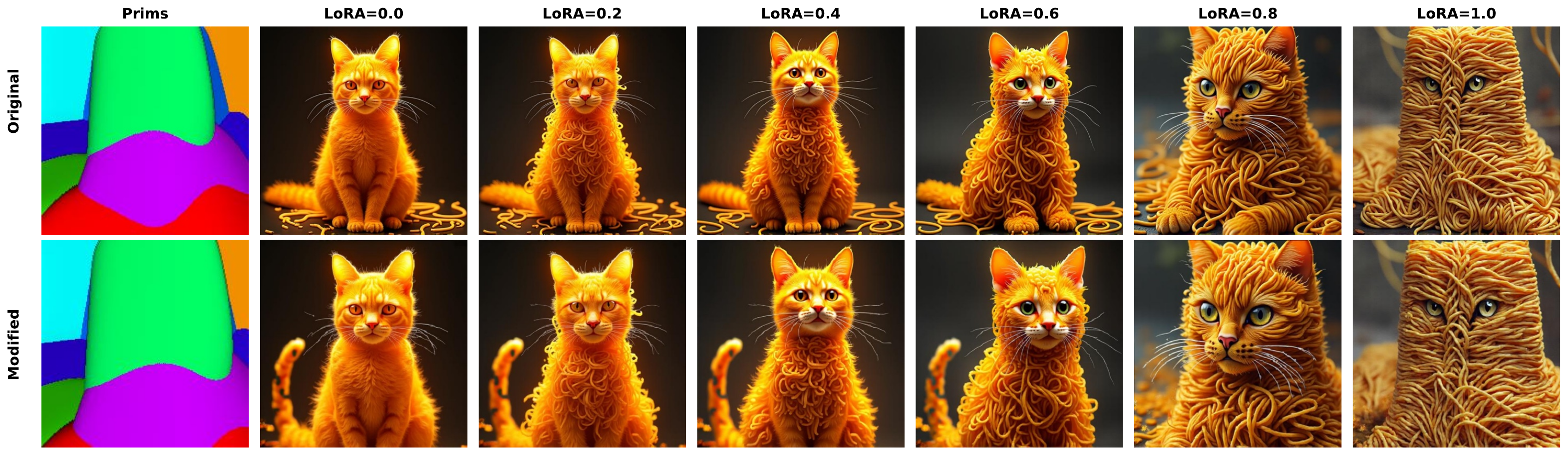}\\[2mm]
    {\fontfamily{pzc}\selectfont ``Cat made of spaghetti''}
\end{minipage}

\vspace{7mm}

\begin{minipage}{\linewidth}
    \centering
    \includegraphics[width=\linewidth]{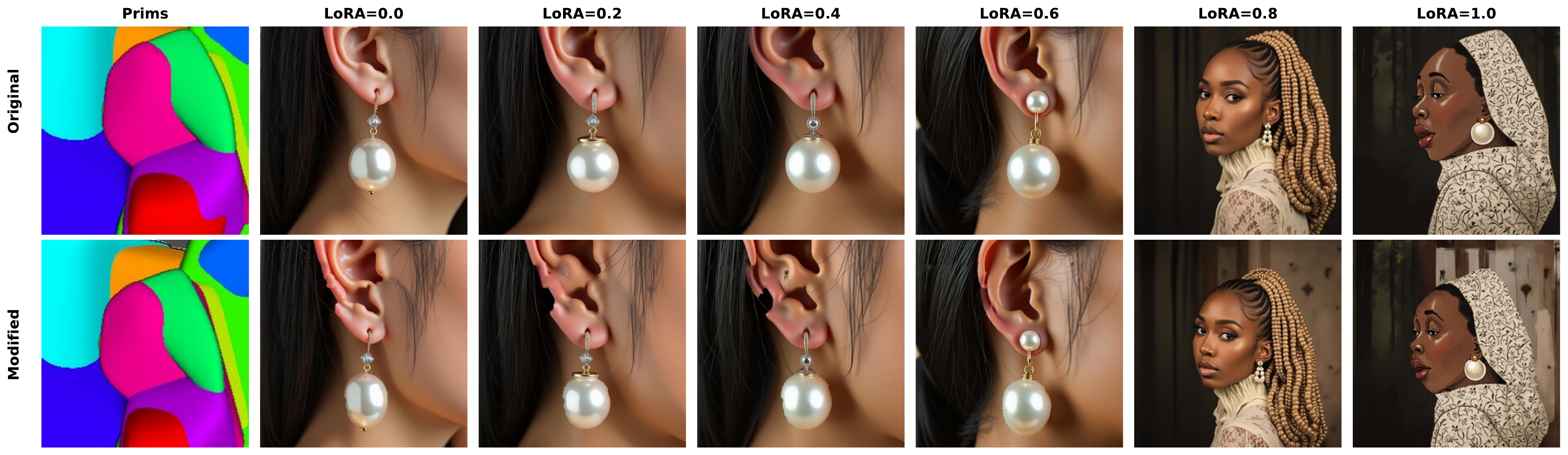}\\[2mm]
    {\fontfamily{pzc}\selectfont ``Girl wearing pearl earrings''}
\end{minipage}

\caption{\textbf{Ablating the effect of $lora_{weight}$ on image editing}. The \textbf{first row} in each set shows source primitives followed by the source (generated) image for $lora_{weight} \in [0.0, 0.2, 0.4, 0.6, 0.8, 1.0]$. As the weight increases, the model more strongly adheres to the depth conditioning. The \textbf{second row} shows edited primitives (first column) followed by the edited results (using our texture-hint mechanism). With weak depth adherence (low LoRA weights), textures do not follow the primitive geometry. Stronger weights (0.8, 1.0) simultaneously preserve both primitives and texture hints.}
\label{fig:EditLoraAblation}
\end{figure*}

\begin{figure}[h]
\centering
\includegraphics[width=\linewidth]{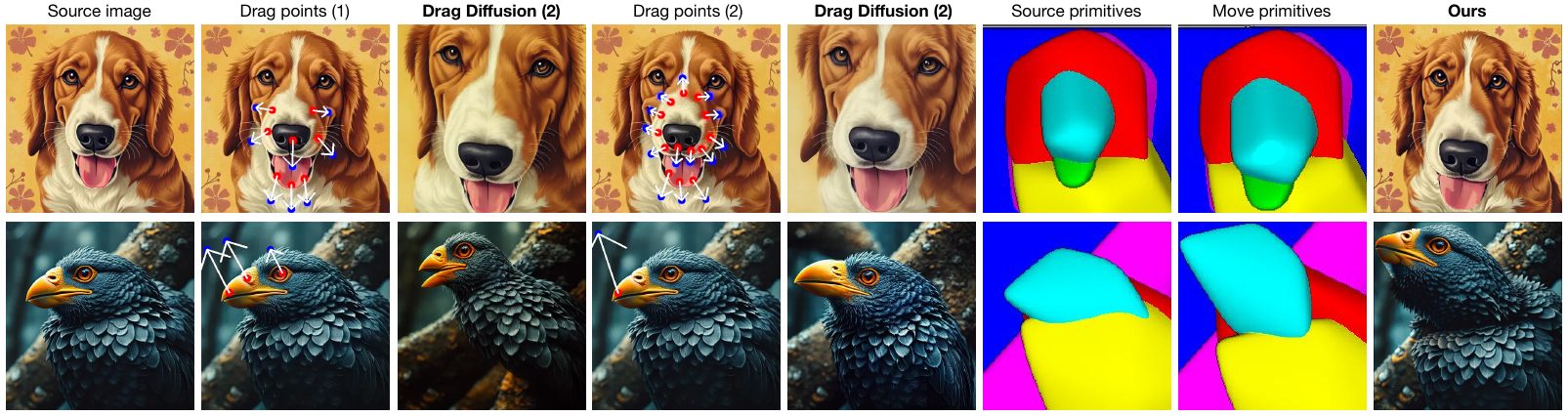}
\caption{\textbf{In addition to Fig.~\ref{fig:edit_1} of the main paper, we provide more comparison with DragDiffusion}~\cite{shi2024dragdiffusion}.  The first column shows the source image. The next two columns show the prompt and result from DragDiffusion. For 3D-aware edits such as enlarging a dog's nose or moving a bird’s head, DragDiffusion’s arrow-based interaction makes precise control difficult; users must effectively draw dense optical-flow vectors. The next two columns show another attempt. DragDiffusion fails to consistently move the bird’s beak to the intended location and alters the texture undesirably. Notice DragDiffusion changes the beak size in both attempts; ours keeps it consistent. It also removes background patterns in the dog example and unintentionally changes the head shape. The final three columns show our approach, which closely follows the primitives while preserving source textures. Our primitive-based interface is more intuitive and produces reasonable part segmentations as a by-product (e.g., red primitive for the dog's head, blue for the snout, green for the tongue).}
\label{fig:drag_2}
\end{figure}

\begin{figure}[h]
\centering
\includegraphics[width=\linewidth]{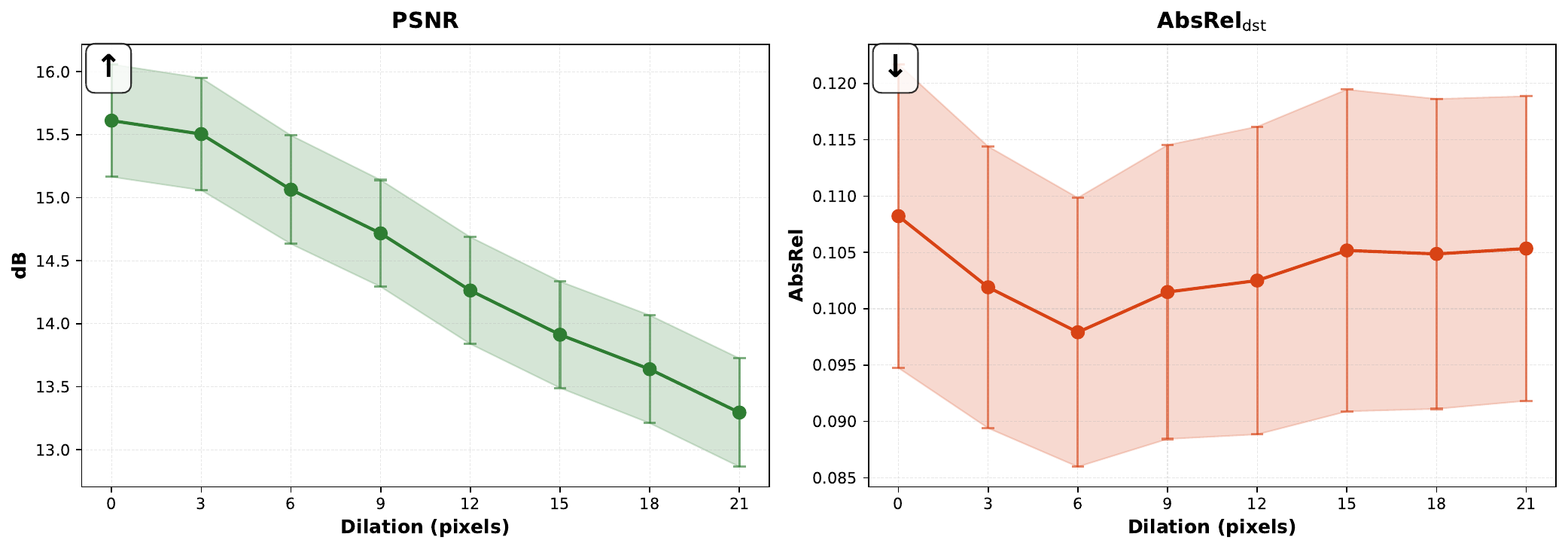}
\caption{\textbf{Quantitative ablation on confidence-mask dilation}. As discussed in Sec.~\ref{sec:hyper}, dilating the confidence mask by a few pixels reduces edge artifacts and improves harmonization of moved objects with the scene. While dilation slightly hurts texture-consistency metrics, it can improve depth accuracy and reduce visible artifacts. Dilating removes warped pixels near primitive boundaries, which often contain misaligned texture due to geometric approximation error. In future work, integrating recent segmentation methods~\cite{Kirillov2023SegmentA} may reduce the need for dilation while maximizing useful hint information. See Fig.~\ref{fig:dilate_tableau} for qualitative examples.}
\label{fig:dilate_analysis}
\end{figure}

\begin{figure*}[h]
\centering
\begin{minipage}{\linewidth}
    \centering
    \includegraphics[width=\linewidth]{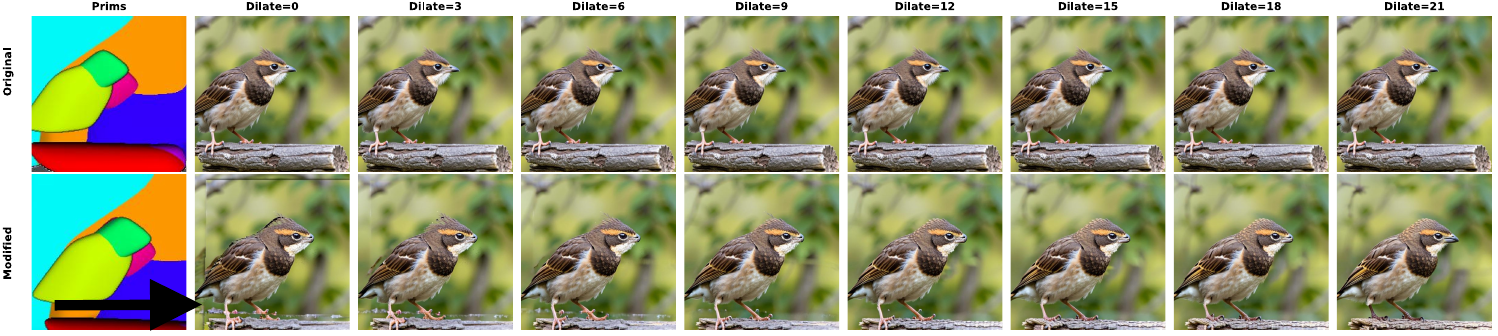}
\end{minipage}

\vspace{7mm}

\begin{minipage}{\linewidth}
    \centering
    \includegraphics[width=\linewidth]{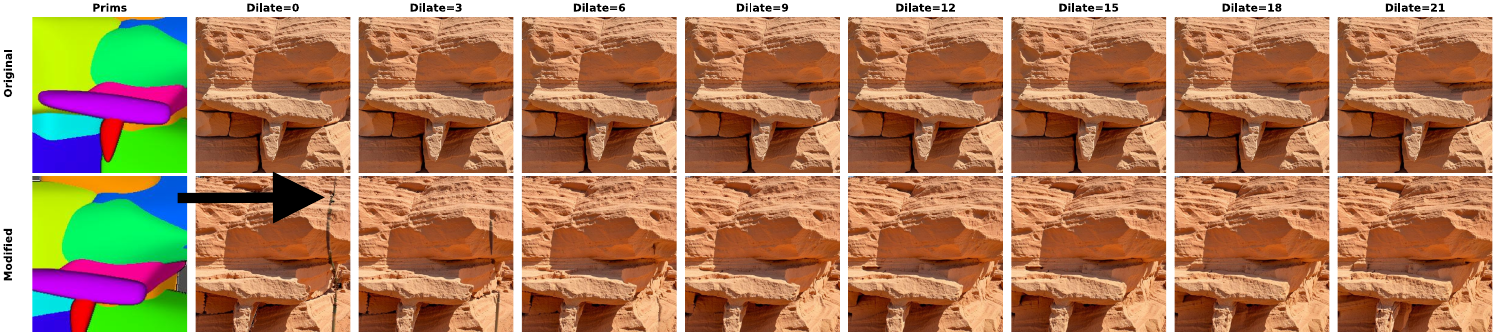}
\end{minipage}

\caption{\textbf{Qualitative ablation on confidence-mask dilation}. The first column in each pair shows the primitives and their edits; subsequent columns show source and edited images across dilation parameters. Dilation expands the unconfident regions of the mask, letting the diffusion model freely synthesize more pixels instead of following the hint. In each pair, the black arrow (bottom left) highlights edge artifacts near hint boundaries. Allowing moderate dilation (typically 9–12 pixels) is sufficient to eliminate these artifacts.}
\label{fig:dilate_tableau}
\end{figure*}

\begin{figure}[h]
\centering
\includegraphics[width=\linewidth]{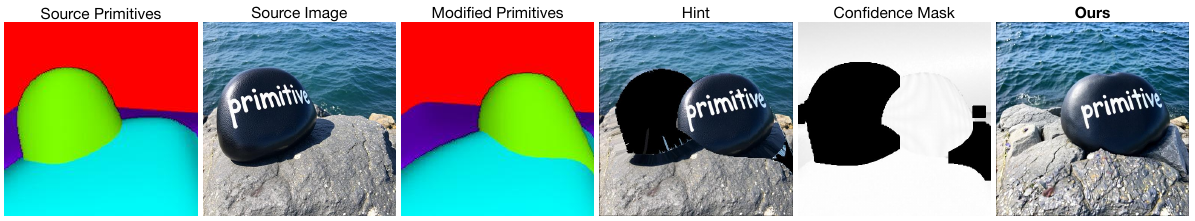}
\caption{\textbf{Move object with hint and confidence mask shown.}}
\label{fig:stone_new}
\end{figure}

\end{document}